\documentclass[12pt,reqno,a4paper]{amsart}
\usepackage{amsmath,amssymb,amsfonts}
\usepackage[mathscr]{eucal}

\newtheorem{proposition}{Proposition}

\newtheorem{corollary}{Corollary}
\newtheorem{lemma}{Lemma}

\theoremstyle{definition}
\newtheorem{remark}{Remark}
\newtheorem{example}{Example}[section]
\newtheorem{definition}{Definition}

\newcommand{\cal}{\EuScript}

\renewcommand{\leq}{\leqslant}
\renewcommand{\geq}{\geqslant}

\renewcommand{\L}{{\cal L}}

\newcommand{\der}[2]{{\frac{\partial {#1}}{\partial {#2}}}}
\newcommand{\lder}[2]{{\partial {#1}/\partial {#2}}}

\newcommand{\RR}{\mathbb R}

\newcommand{\p}{\partial}

\def\vare{\varepsilon}

\def\G {\Gamma}
\def\GG {{\cal G}}
\def\X  {{\bf X}}
\def\Y  {{\bf Y}}
\def\A  {{\bf A}}
\def\F {{\cal F}}
\def\D {{\cal D}}

\def\m{\medskip}

\def\l {{\lambda}}
\def\hl{{\widehat\lambda}}
\def\lo{{\lambda_0}}
\def\hD{\widehat {\Delta}}
\def\wD #1 {{\widehat {\Delta_{#1}}}}
\def\hPi {{\widehat {\Pi}}}

\def\hPrh {{\widehat {P^\rh_\lo}}}

\def\pr{{\rm pr\,}}
\def\ad {{\rm ad}}

\def\Diff {{\rm Diff}}

\def\proj {{\rm proj\,}}
\def\Diff {{\rm Diff\,}}
\def\diff {{\rm diff\,}}
\def\SDiff {{\rm SDiff\,}}
\def\sdiff {{\rm sdiff\,}}
\def\Vol  {{\rm SDiff\,}}
\def\vol  {{\rm sdiff\,}}
\def\uppernabla {{{}^S\nabla}}

\newcommand{\bs}{{\boldsymbol{s}}}

\newcommand{\rh}{{\boldsymbol{\rho}}}

\title[Operator pencil passing through a given operator]
{Operator pencil passing through a given operator}

\author{A.~Biggs}
\author{H.~M.~Khudaverdian}

\address{School of Mathematics,  University of Manchester,
 Oxford Road,  Manchester   M13 9PL,  UK}
\email{khudian@manchester.ac.uk\\adam.biggs@student.manchester.ac.uk }



\begin{document}

\maketitle
\begin{abstract}
Let $\Delta$ be a linear differential operator acting on
the space of densities of a given weight
$\lo$ on a manifold $M$.
One can consider a pencil of operators $\hPi(\Delta)=\{\Delta_\l\}$
passing through the operator $\Delta$ such that any
$\Delta_\l$ is a linear differential operator acting on
denstities of weight $\l$.
This pencil
can be identified with a linear differential
operator $\hD$ acting
on the algebra of densities
of all weights. The existence of an invariant scalar
product in the algebra of densities implies a natural decomposition
of operators, i.e. pencils of
self-adjoint and anti-self-adjoint operators.
  We study lifting maps that are on one hand
equivariant with respect to divergenceless
vector fields, and, on the other hand,
with values in self-adjoint or anti-self-adjoint
operators. In particular we analyze the relation between
these two concepts, and apply it to the study of
$\diff(M)$-equivariant liftings.
%On the base of this analysis we study the problem of existence
%of lifting maps which are equivariant with respect to Lie algebra
%of all vector fields, i.e. all infinitesimal diffeomorphisms.
%We reveal the importance of self-adjointness condition for
%existence of such maps.
Finally we briefly consider the case of liftings
equivariant with respect
to the algebra of projective transformations
and describe all regular self-adjoint and anti-self-adjoint
liftings.

\end{abstract}

 \section {Introduction}

We say that $\bs=s(x)|Dx|^\l$ is a density of weight $\l$
on a manifold $M$ (which we assume to be orientable with a given orientation)
if under a change of local coordinates it is multiplied by the $\l$-th
power of the Jacobian of the coordinate transformation:
           \begin{equation}\label{transformationequations1}
                          \bs
           =s(x)|Dx|^\l=
s\left(x\left(x^\prime\right)\right)\left(\det
 \left({\p x\over \p x'}\right)\right)^\l
                    |D{x^\prime}|^\l\,.
        \end{equation}

   Let $\F_\l=\F_\l(M)$ denote the
 space of
densities of weight $\l$ on $M$ for $\l$ an arbitrary real number.
Note that the space of functions on $M$ is $\F_0(M)$,
denstities of weight $\l=0$.

   Differential operators on the space of densities of
different weights have been under
   intensive study, see
\cite{LecomteMath1}, \cite{ManZag},
\cite{DuvOvs}, \cite{Math2},
\cite{DuvLecomteOvs},
\cite{LecomteOvs1}, \cite{DuvOvs2}, \cite{GarOvsMath}
and the book \cite{OvsTab} and citations therein. See also \cite{KhVor2} and \cite{KhVor4}.

In these works the  spaces $\D_\l(M)$,
  of linear differential operators defined on $\F_\l(M)$
were studied. In particular in
\cite{LecomteMath1}, \cite{DuvOvs} and
\cite{Math2},  the problem of the existence of
$\diff$-equivariant maps  between these spaces
were studied.
The spaces $\D_\l(M)$ can be naturally considered
  as modules not only over group of diffeomorphisms but also
over its subgroups such as
projective or conformal transformations
(provided the manifold is equipped with
  projective or conformal structure).

  In the works \cite{LecomteOvs1}, \cite{DuvLecomteOvs}
the question concerning the existence of a quantisation map
equivariant with respect to
projective or conformal groups was studied.
They constructed a unique full symbol calculus in
each of the above cases, and used this to define the quantisation map.
 On the other hand authors of the works \cite{KhVor2}
and \cite{KhVor4} whilst analysing Batalin-Vilkovisky geometry
and $\Delta$-operators on densities of weight ${1\over 2}$,
naturally came to the analysis of second order operators
acting on the algebra of densities of all weights.
Considering the canonical scalar product in this
algebra, for the classification of second order operators
on odd symplectic supermanifold, they come in particular to
a self-adjoint pencil lifting of second order operators
in the spaces $\D_\l(M)$.
This result gave a clear geometrical picture to the
isomorphisms between modules $\D^{(2)}_\l(M)$
established in \cite{DuvOvs}.

 \subsection{Operator pencils and operators on algebra of densities}
  One can consider pencils of differential operators
(these shall also be referred to as operator pencils), i.e. a family
   $\{\Delta_\l\}$ of operators depending on a parameter $\l$,
such that for each $\l$, $\Delta_\l$ is a differential operator that
   acts on the space $\F_\l$.
  For a simple example of an operator pencil
let $x$ be the standard coordinate  on the line $\RR$ then the formula
   \begin{equation}\label{firstexample}
    \{\Delta_\l\}\colon\qquad \Delta_\l=
A(\l){d^2\over dx^2}+B(\l){d\over dx}+C(\l),\qquad
    \F_\l\to \F_\l\,,
   \end{equation}
   where $A(\l), B(\l), C(\l)$ are functions of $\l$,
defines a pencil of second
   order operators on $\RR$.

   In this paper we shall study pencils of operators passing through a
   given operator which acts on densities of a given weight.
    For further considerations it will be useful to identify pencils
  of operators with operators acting on the algebra of \emph{all} densities,
$\F(M)=\oplus_\l \F_\l(M)$. (If
    $\bs_1=s_1(x)|Dx|^{\l_1}\in \F_{\l_1}$ is a density of weight $\l_1$
    and $\bs_2=s_2(x)|Dx|^{\l_2}\in \F_{\l_2}$ is
a density of weight $\l_2$ then their product is a density
$\bs_1\cdot \bs_2=s_1(x)s_2(x)|Dx|^{\l_1+\l_2}$
    of weight $\l_1+\l_2$. See in more detail \cite{KhVor2}.)

   Consider the linear operator $\hl$ acting on the algebra $\F(M)$
 which multiplies a density by its weight:
   \begin{equation}\label{Euleroperator}
   \hl \bs=\l \bs \quad {\rm if }\,\, \bs \in \F_\l:\quad
   \hl\left(s(x)|Dx|^\l\right)=\l s(x)|Dx|^\l\,.
   \end{equation}
   The operator $\hl$ ({\it weight operator})
 is a first order linear
differential operator on the algebra $\F(M)$ as the Leibnitz rule is obeyed:
      $$
      \hl(\bs_1\cdot\bs_2)=\hl(\bs_1)\cdot\bs_2+\bs_1\cdot(\hl \bs_2)\,.
      $$
Respectively $\hl^n$ is an $n$-th order linear differential operator on $\F(M)$.

This observation allows us to consider operator pencils which depend
 polynomially on $\l$ as differential operators on the algebra $\F(M)$
of all densities. For example the operator pencil
  $\{\Delta_\l\}\colon\,
\Delta_\l=\l{d^2\over dx^2}+(\l^2+1){d\over dx}$,
 on $\RR$ (an operator pencil of the form
 \eqref{firstexample} for
$A(\l)=\l, B(\l)=\l^2+1$, $C(\l)=0$)
 can be considered as a third order operator
                   $$
             \widehat\Delta=
        \hl{\p^2\over \p x^2}+(\hl^2+1){\p\over \p x}
                   $$
on the algebra $\F(\RR)$.
 A general differential operator $\widehat\Delta$ of order $\leq n$
on the algebra $\F(M)$ can be written locally as an expansion of the form
\begin{equation}\label{operatorincoordinates0}
     \widehat\Delta=L^{(n)}+\hl\, L^{(n-1)} +\hl^2\,L^{(n-2)}+
\ldots+ \hl^{n-1}\,L^{(1)}+\hl^n\, L^{(0)}\,,
\end{equation}
where $\hl$ is the weight operator \eqref{Euleroperator},
and the  coefficients  $L^{(i)}$
in the expansion are usual differential operators
of orders $\leq i$ acting on densities.
 The corresponding operator
pencil is
\begin{equation}  \label{operatorincoordinates1}
    \{\Delta_\l\}\colon \,\,
    \Delta_{\l}=\widehat\Delta\big\vert_{\F_\l(M)}=
     L^{(n)}+\l L^{(n-1)}+\l^2 L^{(n-2)}+
  \ldots+ \l^{n-1}L^{(1)}+\l^n L^{(0)}\,.
\end{equation}
It is useful to consider vertical operators and
vertical maps on the algebra of densities:
\begin{definition}
An operator $\hD$ acting on $\F(M)$ is called
vertical if for an arbitrary function $f$ and a density $\bs$
\begin{equation}\label{verticaloperators}
 \hD(f\bs )=f\hD(\bs)\,.
\end{equation}

If $\hD$ is vertical operator,
then $\hD=\sum\hl^k c_k(x)$, and the restriction of
it to any space $\F_\l$ is an operator of order 0,
i.e. multiplication by a function.

Similarly we shall call a map of operators vertical
if its image lies in the space of vertical operators.

\end{definition}

Later by default we shall only consider operator
pencils depending polynomialy on the weight $\l$

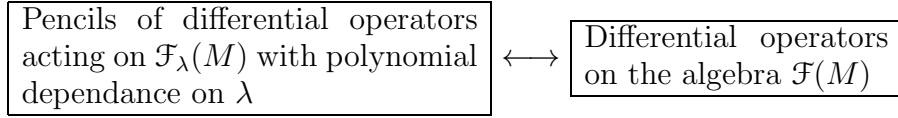
       \centerline {
          \begin{tabular}{|p{6cm}|}
            \hline
 Pencils of differential operators acting on $\F_\l(M)$ with polynomial
dependance on $\l$ \\
   \hline
\end{tabular}
$\longleftrightarrow$
\begin{tabular}{|p{4cm}|}
             \hline
  Differential operators on the algebra $\F(M)$
  \\
  \hline
\end{tabular}
}

\m

   We shall continuously make the identification above throughout the text, namely operator pencils with their
corresponding operators on $\F(M)$.

We shall formulate now explicitly what we mean  when we
say that we ``draw a pencil'' through a given operator.
    Denote by $\D^{(n)}_{\l}(M)$ the space of linear differential
operators of order  $\leq n$
    acting on the space $\F_\l(M)$ and by
    $\widehat{\D}^{(n)}(M)$ the space of linear differential
   operators of order  $\leq n$
    acting on algebra $\F(M)$
    \footnote{An operator $\hD\in \D^{(n)}(M)$
    has an order $\leq n$ if for an arbitrary density $\bs$
     the commutator with the multiplication operator
    $\left[\hD,\bs\right]=\bs\circ\hD-\hD\circ\bs $
 is an operator of order $\leq n-1$.  Respectively
     an operator $\Delta\in \D^{(n)}_\l,\,
\Delta\colon \F_\l\to \F_\l$
    has order $\leq n$ if the operator
    $[\Delta,f]=f\circ\Delta-\Delta\circ f$
    has order $\leq n-1$ for an arbitrary function $f$.}.
   We also consider the spaces $\D_\l(M)=
   \cup_n \D^{(n)}_\l(M)$
of linear differential operators of all orders
  acting
   on $\F_\l$ and respectively the space
   $\widehat{\D}(M)=\cup_n \widehat{\D}^{(n)}(M)$
   of linear differential operators of all orders acting
 on $\F (M)$.

   Recall that an arbitrary linear differential operator
  $\widehat\Delta$ on  $\F(M)$ may be identified with an
   operator pencil $\{\Delta_\l\}$,
    $\hD\big\vert_{\hl=\l}=\Delta_\l$
   ($\hD\big\vert_{\hl=\l}=\widehat\Delta\big\vert_{\F_\l(M)}$)
polynomially depending on $\l$,
thus in particular an operator $\widehat\Delta$
of order $\leq n$ acting on $\F(M)$,
 ($\widehat\Delta\in \widehat{\D}^{(n)}(M)$) can be considered
as an operator pencil, which is a polynomial in $\l$, of order $\leq n$
such that terms which have order $k$ over
the weight operator
$\hl$ ($k\leq n$) possess derivatives
over the coordinates
 $x^i$  of order $\leq n-k$
(see equations \eqref{operatorincoordinates0} and
\eqref{operatorincoordinates1}).

 \begin{definition}
     We say that the operator
 $\hD$ on the algebra of densities is a
pencil lifting of the operator $\Delta$,
(or just a lifting of $\Delta$) if
 the restriction of $\hD$ to the space $\F_\l(M)$
 is the operator $\Delta$ itself:
       \begin{equation*}%\label{defofliftingmap}
        \hD\big\vert_{\hl=\l}=\Delta\,.
         \end{equation*}
 One can say that the operator pencil, which
is identified with the operator
 $\hD$, passes through the operator $\Delta$.

  We consider linear maps $\hPi$ from differential operators on
  densities of an arbitrary but given weight $\l$,
 to operators on the algebra of all densities.
 We say that a map $\hPi$, defined on operators
acting on the space $\F_\l(M)$,
  is a pencil lifting map (or just shortly
lifting map) if the operator $\hD=\hPi(\Delta)$
is a pencil lifting for any operator $\Delta$.
\end{definition}

  \begin{remark}
  In this article we will only consider operators of weight $0$, i.e. operators
  which do not alter the weight of densities.
  (An operator $\Delta$ has weight $\delta$ if $[\hl,\Delta]=\delta\Delta$.)
    In fact there are many interesting
   phenomena associated to operators of non-zero weights
   (see for example the book \cite{OvsTab}
   and citations therein).
   In our analysis these results crop up as well
   and we will consider them later.
   \end{remark}

    \begin{definition}
      Pick an arbitrary $\l_0\in \RR$ and arbitrary $n$.
   Let $\hPi$ be a pencil lifting map defined on the space
     $\D^{(n)}_{\l_0}(M)$:
              \begin{equation*}
\hPi\colon\,\, \D^{(n)}_{\l_0}(M)\rightarrow  \widehat\D(M)\,,
   \,\,\hbox{            }\,\, \hPi\big\vert_{\F_{\l_0}(M)}={\bf id}\,.
     \end{equation*}
  We say that the lifting map $\hPi$ is a {\it regular} lifting  of the space
     $\D^{(n)}_{\l_0}(M)$  if it takes values in operators of
order $\leq n$:
                \begin{equation*}
    \forall \Delta\in \D^{(n)}_{\l_0}(M),\,\,\hPi(\Delta)\in
                \widehat\D^{(n)}(M).
                   \end{equation*}
  We say that $\hPi$ is a {\it strictly regular} lifting map on
     $\D^{(n)}_{\l_0}(M)$  if its restriction on every subspace
    $\D^{(k)}_{\l_0}(M)\subseteq \D^{(n)}_{\l_0}(M)$, ($k\leq n$)
  is a regular lifting map:
                \begin{equation*}
 \forall k\leq n,   \forall \Delta\in \D^{(k)}_{\l_0}(M),\,\,
  \hPi(\Delta)\in
               \widehat\D^{(k)}(M)\,.
                   \end{equation*}
In other words for a given $n$, regular liftings map operators
  of order $\leq n$ to operators of order $\leq n$,
 and strictly regular liftings do not
increase the order of any operator $\Delta$ which has order
less or equal than $n$.
    \end{definition}

\begin{example}
Let $\D^{(2)}_0(\RR^n)$ be the space of second order operators on functions
on Cartesian space $\RR^n$. Consider the map
                   \begin{equation*}
     \hPi(\Delta)=\Delta+\hl \Delta_1=
       S^{ij}(x)\p_i\p_j+A^i(x)\p_i+F(x)+
\hl\left(aS^{ij}(x)\p_i\p_j+b\left(\p_kS^{ki}(x)
    +A^i(x)\right)\p_i\right)\,,
                   \end{equation*}
with values in operator pencils,
operators on the space of all densities on $\RR^n$,
where $a$ and $b$ are arbitrary parameters.
 This is a lifting map since
$\widehat\Delta=\Pi(\Delta)\big\vert_{\l=0}=\Delta$ and assigns
to the second order operator $\Delta=S^{ij}\p_i\p_j+A^i\p_i+F$
the pencil $\{\Delta_\l\}$ of second order operators,
$\Delta+\l\Delta_1$
passing through the operator $\Delta$.
On the other hand this pencil defines a
{\it third} order operator on the algebra of densities in
the case if $a\not=0$ and $S^{ij}\not=0$, since the operator
          $
 \hl\Delta_1=a\hl S^{ij}\p_i\p_j+\dots
           $ acting on $\F(\RR^{n})$
has order three.
 We see that the lifting
map $\Pi(\Delta)$ is a regular map if and only if $a=0$.
This regular map is strictly regular if and only if $b=0$ also,
since in the case $\Delta$ is a first order operator and
 $b\not=0$, the operator
$\Pi(\Delta)=\hl bA^i\p_i$ is a second order operator.
(We will give another more
geometrical examples later in the text.)

 \end{example}

In what follows we consider liftings maps
that are equivariant with respect to
some subgroup of the group of diffeomorphisms or with respect to
some subalgebra of the algebra of vector
fields (infinitesimal diffeomorphisms).

 A lifting map, $\Delta\mapsto \hPi(\Delta)$,
is equivariant with respect to
the diffeomorphism $\varphi$ if $\varphi^*\left(\hPi(\Delta)\right)=
  \hPi\left(\varphi^*\Delta\right)$
for an arbitrary operator $\Delta$.  Respectively, a
lifting $\Delta\mapsto \hPi(\Delta)$ is
equivariant with respect to the vector field $\X$ on $M$
(an infinitesimal diffeomorphism)
if
  \begin{equation*}\label{definitionofinvariance}
\ad_\X\hPi(\Delta)=\hPi\left(\ad_\X \Delta\right)\,,
\end{equation*}
for an arbitrary operator $\Delta$,
 where we denote by $\ad_\X$ the action of $\X$ on an operator
$\Delta$: for an arbitrary density $\Psi$
  \begin{equation*}\label{definitionofinvariance2}
\ad_\X \Delta(\Psi)=
\L_\X\left(\Delta\left(\Psi\right)\right)
-\Delta\left(\L_\X(\Psi)\right)\,,
\end{equation*}
where $\L_\X$ is the Lie
derivative with respect to vector field $\X$.

\begin{definition}
Let $G\subseteq \Diff(M)$ be a subgroup and let
$\GG=\GG(G)$ be the corresponding Lie subalgebra of vector fields.
A pencil lifting map
$\hPi$ is $G$-equivariant
if it is equivariant with respect to
all $\varphi\in G$.
Respectively a lifting map
$\hPi$ is $\cal G$-equivariant if it is
equivariant with respect to an arbitrary vector field $\X\in {\cal G}$.
(We will also refer to such maps as
$G$-liftings respectively $\GG$-liftings.)
\end{definition}

    In this article we primarily consider $\SDiff\,_{\rh}(M)$,
    that is the group of diffeomorphisms that preserve a volume $\rh$ on
    $M$, and the corresponding algebra, $\sdiff\,_{\rh}(M)$, of
    divergenceless vector fields:
              \begin{equation}\label{divergenceless}
\X\in \vol_\rh\colon \quad {\rm div}_\rh\,\X=
\rh^{-1}\L_\X\rh=
\p_iX^i(x)+X^i(x)\p_i \log\rho(x)=0\,,
\,\quad (\rh=\rho(x)|Dx|)\,.
              \end{equation}
We shall also discuss
    equivariance with respect to the full group of
    diffeomorphisms, and in the final section we shall
    consider projective transformations.

Technically it is more convenient to consider invariance with
respect to a Lie algebra rather than the group. We will mostly
consider  liftings equivariant with respect to a Lie algebra.

 On  $\F(M)$ one can consider the
      canonical scalar product, $\langle\,\,,\,\,\rangle$,
defined by the following formula:
  for $\bs_1=s_1(x)|D{x}|^{\l_1}$ and $\bs_2=s_2(x)|D{x}|^{\l_2}$ then
                 \begin{equation}\label{canonicalscalarproduct}
                   \langle \bs_1,\bs_2\rangle
                          =
                      \begin{cases}
                        & \int_M s_1(x)s_2(x)|D{x}|\,,
                 \quad {\rm if}\,\l_1+\l_2=1\,,\cr
                                                               \cr
                        &  0\,\, {\rm if}\,\,
        \quad {\rm if}\,\l_1+\l_2\not=1\,.\cr
                      \end{cases}
                 \end{equation}
This implies that a linear differential
operator $\widehat \Delta$
acting on
the algebra $\F(M)$ has an adjoint operator $\widehat\Delta^*$:
          \begin{equation}\label{canonicalscalarproduct1}
           \langle \Delta \bs_1,\bs_2\rangle=
             \langle \bs_1,\Delta^*\bs_2\rangle\quad
           \left(\widehat\Delta^*\right)\big\vert_{\hl=\l}=
        \left(\hD\big\vert_{\hl=1-\l}\right)^*\,.
           \end{equation}
One can see that $\hl^*=1-\hl$,
$\left(\p/\p x^i\right)^*=-\left(\p/\p x^i\right)$.
(For details see \cite{KhVor2} or \cite{KhVor4}.).
 We thus come to the notion of a self-adjoint
(anti-self-adjoint) operator
on the algebra of densities.
These constructions, which were
introduced in \cite{KhVor2}
for analysis of second order operators, will be in active use
in our considerations.
In particular we can consider self-adjoint and anti-self-adjoint
lifting maps, where a lifting map, $\hPi$, is called self-adjoint
(anti-self-adjoint) if
for an arbitrary operator $\Delta$ the operator
$\hPi(\Delta)$ is a self-adjoint (anti-self-adjoint) operator
on the algebra of densities.

\begin{remark}\label{why operator is self-adjoint?}
If $\Delta=\hl^rS^{i_1\dots i_{n-r}}
\p_{i_1}\dots\p_{i_{n-r}}+\ldots$
is an $n$-th order operator such that the tensor $S^{i_1\dots i_{n-r}}$
 does not vanish identically, then

$$\Delta^*=({\hl^*})^r(-1)^{n-r}S^{i_1\dots i_{n-r}}
\p_{i_1}\dots\p_{i_{n-r}}+\ldots=$$
        \begin{equation*}\label{whyself-adjointorantiselfadjoint}
       ({1-\hl})^r(-1)^{n-r}S^{i_1\dots i_{n-r}}
\p_{i_1}\dots\p_{i_{n-r}}+\ldots=
   (-1)^n\hl^rS^{i_1\dots i_{n-r}}
\p_{i_1}\dots\p_{i_{n-r}}+\ldots\,,
 \end{equation*}
i.e. the operator  $\Delta - (-1)^n\Delta^*$
is an operator of the order $\leq n- 1$.

In particular it follows from this fact that if
we consider liftings of operators
defined on the spaces  $\D^{(n)}_\lo$
(operators of the order $\leq n$ on densities of weight $\lo$),
then self-adjoint pencil liftings have to be considered
if $n$ is an even number, and
anti-self-adjoint pencil liftings
if $n$ is an odd number.

\end{remark}

  This article contains the following:

  In section \ref{section2} we recall
constructions of canonical pencils on the spaces
  of first order and second order operators. Pencil liftings for
    first order operators is just a
reformulation of standard exercises in differential geometry
     concerning Lie derivatives of functions and densities.
For second order operators we briefly recall the constructions
suggested in \cite{KhVor2}, where a canonical self-adjoint pencil
    of second order operators was constructed.
It turns out that for every second order
    operator $\Delta\in \D_\l(M)$ ($\l\not=0,1/2,1$)
there exists a unique such a pencil.
    This pencil provides us with a regular
$\diff(M)$-equivariant pencil lifting map on $\D^{(2)}_\l(M)$.
    Restricting this pencil  for different
values of $\l^\prime$ leads to isomorphisms
    of $\diff$-modules $\D^{(2)}_\l$ and
$\D^{(2)}_{\l^\prime}$ ($\l,\l^\prime\not=0,1/2,1$).
    These are just the isomorphisms established in the
work \cite{DuvOvs} of Duval and Ovsienko.
    The canonical self-adjoint pencil reveals not only
the geometrical meaning of
    their isomorphisms but also indicates the special
role of self-adjointness for maps between spaces
    of operators.

    What about maps between spaces of higher order operators?
     Results of the works \cite{LecomteMath1} and \cite{Math2}
imply sort of no go theorem that there
      do not exist $\diff(M)$-equivariant liftings on
     $\D^{(n)}_\l(M)$, if manifold $M$ has
dimension greater than 1.
         We shall attempt to analyze the geometrical
reasons behind this.
         In section \ref{section3} and \ref{section4}
         we consider a smaller algebra
         of divergenceless vector fields, $\sdiff(M)$,
          corresponding to the group $\SDiff(M)$ of diffeomorphisms
          preserving  a volume form.
 On one hand the volume form
         identifies all the spaces $\F_\l(M)$ and hence the spaces
         of operators $\D_\l(M)$.
         On the other hand any candidate for
      $\diff(M)$-equivariant map
         has to pass the test of
        being at least $\sdiff(M)$-equivariant.
 We describe all regular $\sdiff$-equivariant
         pencil liftings
  and (anti)-self-adjoint regular $\sdiff$-pencil liftings
if manifold $M$ has dimension greater than 2.
         It turns out that only self-adjoint
     $\sdiff_\rh(M)$-liftings have a chance to be
         $\diff(M)$-equivariant. Namely  we see how a
         pencil lifting depends on a volume form,
        and come to the conclusion that if
         $\sdiff_\rh$-lifting map does not depend on a
           volume form $\rh$ (i.e. it is
          $\diff(M)$-lifting) at least in a vicinity of a given operator
         then the map is (anti)-self-adjoint
       in a vicinity of this point (up to a vertical operator)
         (see corollary \ref{corollary1} below).

        In the section \ref{section5}
we suggest Taylor series expansion of operators
        on algebra of densities with
respect to vertical operators.
         This Taylor
series expansion may be useful for different tasks. In particular
        we use this expansion to describe all (anti)-self-adjoint liftings
        of an arbitrary operator acting on densities.

    Finally in
section \ref{section6} we briefly consider regular pencil liftings
    which are
equivariant with respect to the algebra of projective transformations.
    We write down the formula for all regular liftings equivariant
    with respect to infinitesimal projective transformations of $\RR^d$
    and calculate those that are
    (anti)-self-adjoint amongst them.

\begin{remark} We only consider linear maps on operators
that are differential polynomials. These maps are non-other
than those that are
local by Peetre's theorem \cite{Peetre}.
(In fact in the case of equivariant maps
it can be shown that for suitable algebras of vector
fields this is automatically
satisfied (details can be found in
\cite{LecomteMath1} or \cite{LecomteOvs1}.))

  In this article by default the manifold $M$ under consideration
  is oriented
and therefore the transformation laws \eqref{transformationequations1}
are well defined for  {\it arbitrary} $\l$.  For the
adjoint of an operator  to be well defined
  we suppose that $M$ is compact, or it is
  an open domain in $\RR^m$. In the latter case we assume that the
functions are rapidly decreasing at infinity.

\end{remark}

\textbf{Acknowledgement.}  We are very grateful to
V. Ovsienko and Th. Voronov for many
encouraging discussions and advice throughout
the time which we were working on
this article. We would also like to send our thanks
to P. Mathonet  and F. Radoux for their stimulating discussions.
One of us (H.M.Kh.) is very happy to acknowledge the
wonderful environment of the MPI Bonn
and I.H.E.S Paris, which was very helpful during the
first phase of the work on this paper.

\section {Canonical liftings of first and
second order operators}\label{section2}

In this section we consider canonical constructions that give us liftings
of first and second order operators which are
$\Diff(M)$-equivariant. In subsequent sections
we will see that all
$\Diff(M)$-liftings
that are defined on all operators of a given weight and order
are contained in these examples.

\subsection{Lifting of first order operators}\label{subsection2.1}

A vector field $\X$ on $M$
defines a Lie derivative
$\L_\X^{\l}$, a first order operator
on the space $\F_\l(M)$ of
densities of weight $\l$.
   If in local coordinates  $(x^i)$,
  $\X=X^i(x)\p_i$ ($\p_i={\p\over \p x^i}$)
then
                 \begin{equation*}\label{liederivative1}
              \L^{\l}_\X(\bs(x))=
    \L^{\l}_\X\left(s(x)|Dx|^\l\right)=
                     \left(
   X^i(x)\p_is(x)+\l\p_i X^i(x)s(x)
                    \right)|Dx|^\l\,.
               \end{equation*}
A pencil of Lie derivatives $\{\L_\X^{\l}\}$ can be identified
with the Lie derivative $\widehat\L_\X$, a first order operator
on $\F(M)$:
                 \begin{equation}\label{liederivative2}
\widehat\L_\X=
   X^i(x)\p_is(x)+\hl\p_i X^i(x)\,,\quad
   \left(\widehat\L_\X\right)\big\vert_{\hl=\l}=\L^{\l}_\X\,.
 \end{equation}
This is an anti-self-adjoint operator:
                 \begin{equation*}\label{liederivative3}
\left(\widehat\L_\X\right)^*=
                   \left(
X^i(x)\p_i+\hl\p_iX^i
                  \right)^*
                        =
-X^i(x)\p_i-\p_i X^i(x)
   +\hl^*\p_i X^i(x)=
                 -\widehat\L_\X
 \end{equation*}
since $\hl^*=1-\hl$.
In other words the operator
 $(\L^\l_\X)^*$ acting on densities of weight ($1-\l$) is equal to
$-\L^{1-\l}_\X$.

Pick any $\lo$ and consider a space
$\D^{(1)}_{\l_0}(M)$ of first order operators acting on
densities of weight $\lo$. For an arbitrary operator
$\Delta\in\D^{(1)}_{\lo}(M)$
consider the vector field $\A$,
the principal symbol of the
operator $\Delta$:
$\left(\A f\right)\bs=\Delta\left(f\bs\right)-f\Delta(\bs)$,
where $f$ is an arbitrary function and
 $\bs$ is an arbitrary density of weight $\l_0$ on $M$.
The difference between the operator $\Delta$ and the Lie derivative
$\L^\lo_\A$
is a zeroth order
  operator.
We thus come to the conclusion that
an arbitrary first order operator $\Delta$
can be canonically decomposed into
the sum of a Lie derivative and a scalar function:

    $$  \Delta=
A^i(x)\p_i+B(x)=\L_\A^{(\l_{0})}+S(x)\,,
    $$
 \begin{equation}
    \label{geometricalmeaningoffirstorderoperator}
{\rm where}\qquad S(x)=\Delta-\L_\A^{(\l_{0})}=
  B(x)-\l_0\p_i A^i(x)\,.
                   \end{equation}
Using this decomposition
we come to a canonical pencil lifting map defined on
the space $\D^{(1)}_\lo(M)$:
          \begin{equation*}
    \label{...}
      \hPi(\Delta)=\widehat\L_\A+S(x)\,,\quad {\rm if}
    \,\,\Delta= \L_\A^\lo+S(x)\,,
           \end{equation*}
where $\widehat\L_\A$ is the Lie derivative \eqref{liederivative2}.
In local coordinates
          \begin{equation}
      \label{canonicalliftingoffirstorderoperator1}
\hPi(\Delta)=\hPi\left(A^i\p_i+B(x)\right)
                   =
          \underbrace
{A^i(x)\p_i+\hl\p_i A^i(x)}_
{\widehat\L_\A}
       +
      \underbrace
  {B(x)-\lo\p_i A^i(x)}_
     {\hbox {scalar function}}\,.
         \end{equation}
This is a pencil lifting map,
$\hPi(\Delta)\big\vert_{\hl=\lo}=\Delta$,
it is strictly regular and it is obviously a $\Diff(M)$-lifting
since the map is canonical, it
does not depend on a choice of local coordinates.

One can twist this lifting to get a family of regular pencil liftings:
 $\Pi(\Delta)=\widehat\L_\A+C(\hl)S(x)$, where $C(\hl)$
is a polynomial of order $\leq 1$
such that it obeys the condition $C(\hl)\big\vert_{\hl=\lo}=1$,
i.e. $C(\hl)=1+c(\hl-\lo)$, where
$c$ is an arbitrary constant.
  We come to the following regular pencil liftings of $\D^{(1)}_\lo(M)$:
          $$
\hPi_c(\Delta)=\widehat\L_\A+\left(1+c(\hl-\lo)\right)S(x)=
           $$
          \begin{equation}
    \label{canonicalliftingoffirstorderoperator2}
           A^i(x)\p_i+\hl \p_i A^i(x)
       +\left(1+c(\hl-\lo)\right)
   \left(B(x)-\lo\p_iA^i(x)\right)\,,
\,\, c\in \RR\,.
           \end{equation}
This formula presents a one-parametric family, an affine line, of
$\Diff(M)$-equivariant
regular pencil liftings. This affine line
possesses the distinguished point
$c=0$, the unique point that corresponds to a strictly regular pencil
\eqref{canonicalliftingoffirstorderoperator1}.
All other points on the line
are not strictly regular liftings
(if $\A\equiv 0$ and $B(x)\not=0$,
then
$\Delta$ is a zeroth order operator while
$\Pi(\Delta)$ is a first
order operator).

In the case if our original weight
$\lo\not={1\over 2}$
the affine line of liftings maps possesses
another distinguished point:
if $c={2\over 2\lo-1}$ then the lifting
\eqref{canonicalliftingoffirstorderoperator2}
becomes
         \begin{equation}
    \label{canonicalliftingoffirstorderoperator3}
\hPi(\Delta)=\widehat\L_\A+{2\hl-1\over 2\lo-1}S(x)
                   =
           A^i(x)\p_i+\hl\p_i A^i(x)
       +{2\hl-1\over 2\lo-1}
   \left(B(x)-\lo\p_i A^i(x)\right)\,.
         \end{equation}
This is an anti-self-adjoint lifting:
$\left(\hPi\left(\Delta\right)\right)^*=-\hPi\left(\Delta\right)$
since $\widehat \L_\A^*=-\widehat \L_\A$
and $(2\hl-1)^*=(2(1-\hl)-1)=-(2\hl-1)$.

\begin{remark} Note that for an arbitrary first order operator
$\hD=M^i(x)\p_i+\hl N(x)+P(x)$ the condition
of anti-self-adjointness,
$\Delta^*=
-\Delta$, is equivalent to the condition that
$N=\p_iM^i(x)-2P(x)$, i.e.
       \begin{equation*}\label{conditionofantislefadjointness}
\Delta=M^i(x)\p_i+\hl\p_i M(x)
   -(2\hl-1)P(x)=\widehat\L_{\bf M}-(2\hl-1)P(x)\,.
       \end{equation*}
Comparing with equation \eqref{canonicalliftingoffirstorderoperator3}
we see that in the case $\lo\not={1\over 2}$ there is a unique
anti-self-adjoint regular lifting map defined on the space
 $\D^{(1)}_\lo(M)$.
\end{remark}

\subsection{Liftings of second order operators}\label{subsection2.2}
We described liftings of first order operators using Lie derivatives
\eqref{liederivative2},  which are  anti-self-adjoint
first order operators. To study canonical liftings of second order operators it is useful
to recall the description of self-adjoint second order operators
(for details see \cite{KhVor2} or \cite{KhVor4}).
Let $\hD$ be an arbitrary second order operator
on $\F(M)$:
   \begin{equation*}\label{secondordergeneral}
   \hD=
    \underbrace{S^{ij}(x)\p_i\p_j+
   \hl B^i(x)\p_i+\hl^2
    C(x)}_{\hbox{second order derivatives}}+
\underbrace{D^i(x)\p_i+
  \hl E(x)}_{\hbox{first order derivatives}}+
          F(x)\,.
           \end{equation*}
The adjoint of this operator has the form:
                $$
                \Delta^*=
            S^{ij}(x)\p_i\p_j+
       2\p_i S^{ij}(x)+
         \p_i\p_jS^{ij}(x)+
                     $$
                     $$
            \left(\hl-1\right)\left(B^i(x)\p_i+
                 \p_i B^i(x)\right)+
            \left(\hl-1\right)^2C(x)-
            \left(\hl-1\right)E(x)-D^i\p_i-
            \p_i D^i(x)
                  +F(x)\,.
                $$
We see that the condition $\Delta^*=\Delta$ is equivalent to
             \begin{equation}\label{selfadjointoperator0}
   \hD=
        S^{ij}(x)\p_i\p_j+
 \p_jS^{ji}(x)\p_i+
 \left(2\hl-1\right) \gamma^i(x)\p_i+
   \hl\p_i\gamma^i(x)+
         \hl
 \left(\hl-1\right)\theta(x)+F(x)
                   \,.
           \end{equation}
Here we denote  $\gamma^i(x)=B^i(x)/2$ and $\theta(x)=C(x)$.
The coefficients of the operator \eqref{selfadjointoperator0}
have the following geometrical meaning:
\begin{itemize}
    \item   $S^{ij}$ is a symmetric
    contravariant tensor field,
        \item   $\gamma^i$ is an upper connection
    (see appendix \ref{appendix1}),

\item  $\theta(x)$ is a Branse-Dicke function,
(see appendix \ref{appendix1}),

\item $F(x)$ is a scalar function. Usually we enforce
the normalisation condition
          \begin{equation}\label{normalisationcondition2}
     F(x)=\hD(1)=0\,.
 \end{equation}

\end{itemize}

Pick an arbitrary $\lo$ and let $\Delta=S^{ij}(x)\p_i\p_j+T^i(x)\p_i+R(x)
\in \D^{(2)}_\lo(M)$. The self-adjoint operator $\hD$ defined by
equation \eqref{selfadjointoperator0}
is a lifting of the operator $\Delta$, i.e.
$\hD\big\vert_{\hl=\l_0}=\Delta$,
if the following conditions hold:
     \begin{equation}\label{condition2}
T^i(x)=\p_jS^{ji}(x)+(2\lo-1)\gamma^i(x)\,,\quad
 R(x)=\lo\p_i\gamma^i(x)+
  \lo(\lo-1)\theta(x)+F(x)\,.
     \end{equation}
We come to the statement that if $\l_0\not=0,{1\over 2}, 1$ then
for every operator  $\Delta\in \D^{(2)}_\lo(M)$,
there exists a unique second order self-adjoint
operator pencil, $\hD$, which
obeys the normalisation condition \eqref{normalisationcondition2},
and passes through this operator $\Delta$.
This pencil is defined by equations \eqref{selfadjointoperator0},
\eqref{condition2} and \eqref{normalisationcondition2}.
  In other words an arbitrary second order operator
$\Delta\in \D^{(2)}_\lo(M)$ ($\lo\not=0,{1\over 2},1)$
uniquely defines the geometrical data $(S^{ik},\gamma^i,\theta)$
(a symmetric contravariant tensor field of rank 2, an upper connection and a
Branse-Dicke function),
and these geometrical objects uniquely
define the self-adjoint second order
operator \eqref{selfadjointoperator0}
on $\D^{(2)}_\lo(M)$
if we enforce normalisation
condition \eqref{normalisationcondition2}.
(For more detail see \cite{KhVor2}, \cite{KhVor4}.)
This canonical
map does not depend on a choice
of local coordinates.
Thus we have defined a regular, self-adjoint,
$\diff(M)$-equivariant pencil lifting map
on the space $\D^{(2)}_\lo(M)$
for $\lo\not=0,{1\over 2},1$.

  Do there exist any other regular $\diff(M)$-liftings?
Uniqueness of this pencil map on $\D^{(2)}_\lo(M)$
follows from a theorem proved by Duval and
Ovsienko in \cite{DuvOvs}.
In the following sections we will give an
alternative proof of this result when analysing $\diff(M)$-liftings
for higher order operators
(see section \ref{subsection4.3}).

\section {$\Diff(M)$ and $\Vol(M)$-equivariant liftings}\label{section3}

In this section we consider $\Vol(M)$-equivariant regular
lifting maps,
which we  will then use
for analysing regular $\Diff(M)$-lifting maps.

Let $M$ be an orientable compact manifold provided with an orientation.
We say that $M$ has a volume form structure if
a volume form $\rh$ is defined on $M$.
Consider the group $\Vol_\rh$ of orientation preserving
 diffeomorphisms which preserve volume form
$\rh$, and its
corresponding Lie algebra
$\vol_\rh(M)$ of divergence-less vector fields (see condition \eqref{divergenceless}).

On a manifold with a volume form $\rh=\rho(x)|Dx|$
an arbitrary density $\bs_1=s_1|Dx|^\l_1$ of weight $\l_1$
can be canonically identified
with a density
$\bs_2=\bs_1\rh^{\l_2-\l_1}=
s_1(x)\rho(x)^{\l_2-\l_1}|Dx|^{\l_2}$
of weight $\l_2$. Thus the two spaces $\F_{\l_1}(M)$
and $\F_{\l_2}(M)$ of densities of weights $\l_1$ and $\l_2$
are naturally identified. This implies
canonical isomorphisms between differential operators on the spaces of densities:
       \begin{equation}\label{identification1}
P_{\l_2,\l_1}^\rh\colon\quad \D_{\l_1}(M)\rightarrow \D_{\l_2}(M),\qquad
 P_{\l_2,\l_1}^\rh\left(\Delta_{\l_1}\right)=
\rh^{\l_2-\l_1}\circ\Delta_{\l_1}\circ\rh^{\l_1-\l_2}\,.
\end{equation}

 The isomorphisms $P^\rh_{\l_1,\l_2}$ define
lifting maps on the spaces $\D_\l(M)$.
Namely choose an arbitrary $\lo$ and assign to every operator
$\Delta_\lo$ on densities of weight $\lo$, a pencil of operators
$\{\Delta_\l\}$ such that $\Delta_\l=P^\rh_{\l,\lo}(\Delta_\lo)$.
This pencil can be identified with an operator
on the algebra of densities:
     \begin{equation}\label{canonicalpencillifting1}
\widehat\Delta=\widehat P^\rh_{\lo}(\Delta_\lo)\colon\,\,
\widehat\Delta\big\vert_{\hl=\l}=P^\rh_{\l,\lo}(\Delta_\lo)\,.
  \end{equation}
This formula defines a natural lifting $\widehat {P^\rh_\lo}$  on the
space $\D_\lo(M)$.
In local coordinates, $(x^i)$, it has the following appearance:
                     $$
  \Delta_\lo=\sum L^{i_1\dots i_k}(x)\p_{i_1}
     \dots \p_{i_k}\mapsto  \hD=
            \widehat {P^\rh_\lo}
         \left(\Delta_\lo\right)=
       \rh^{\hl-\lo}\circ \Delta\circ\rh^{\lo-\hl}=
                 $$
                 $$
        \rho^{\hl-\lo}(x)\circ
  \left(\sum L^{i_1\dots i_k}(x)\p_{i_1}
       \dots \p_{i_k}\right)
           \circ\rho^{\lo-\hl}(x)=
                $$
\begin{equation}\label{localexpression1}
          \sum L^{i_1\dots i_k}(x)
         \left(\p_{i_1}+(\hl-\lo)\G_{i_1}(x)\right)\cdots
         \left(\p_{i_k}+
(\hl-\lo)\G_{i_k}(x)\right)\,.
            \end{equation}
Here $\rh=\rho(x)|Dx|$ is the volume form in coordinates $x^i$ and
$\Gamma_i=-\p_i\left (\log \rho(x)\right)$ is a flat connection
on densities of weight $\l=1$ corresponding
to the volume form $\rh=\rho(x)|Dx|$.
  In brief we come to the canonical pencil lifting
  $\hD=\widehat P^\rh_\lo(\Delta_\lo)$ by changing partial derivatives
$\p_i$ for the covariant derivatives $\nabla_i=\p_i+(\hl-\lo)\G_i$.
The operator $\hD$ is an
operator on the algebra of densities of the same order
 as the operator $\Delta_\lo$.

 \begin{remark}
  The above expression still makes
sense when $\G_i$ is an arbitrary connection.
  We shall only consider the above
case when $\G_i$ corresponds to a volume form.
  (Note that in the case $\G_i$ is
flat the map \eqref{localexpression1} is not just a linear map
  but a map of algebras.)
  \end{remark}

It is important to note that the
 adjointness operation \eqref{canonicalscalarproduct1}
commutes
with the identification isomorphisms $P^\rh_{\l_1,\l_2}$:
            \begin{equation*}\label{adjointforcanonical1}
\forall\,\, \Delta\in \D_\lo(M)\qquad
        \left( P^\rh_{\l,\lo}\left(\Delta\right)\right)^*=
            P^\rh_{1-\l,1-\lo}\left(\Delta^*\right)\,,
         \end{equation*}
and for the canonical lifting \eqref{canonicalpencillifting1}
            \begin{equation}\label{adjointforcanonical2}
            \forall\,\, \Delta\in \D_\lo(M)\qquad
       \left( \widehat{P^\rh_{\lo}}\left(\Delta\right)\right)^*=
           \widehat {P^\rh_{1-\lo}}\left(\Delta^*\right)\,.
         \end{equation}
(Recall that $\Delta^*\in\D_{1-\l}(M)$ if $\Delta\in\D_{\l}(M)$. )

On manifolds with a volume form structure
using the indentification isomorphisms
$P^\rh_{\l_1,\l_2}$ one can consider
not only the
canoncial  scalar product \eqref{canonicalscalarproduct}
on $\F(M)$, but also the
 scalar product defined on densities
of fixed weights: for $\bs_1,\bs_2\in\F_\l(M)$,
$\langle\bs_1,\bs_2\rangle_\rh=\int \bs_1\cdot \bs_2\cdot \rh^{1-2\l}$.
Similarly for an operator $\Delta\in\D_\l(M)$
one can consider its adjoint
(with respect to the volume form
$\rh$),
 such that it acts on densities of weight $\l$ also:
$\Delta^{*_\rh}=P^\rh_{\l,1-\l}(\Delta^*)$. In particular if
an operator $\Delta$ acts on usual functions then its
canonical adjoint \eqref{canonicalscalarproduct1}
$\Delta^*$
acts on densities of weight $1$ whilst its adjoint with
respect to the volume form $\rh$ is
an operator acting on functions:
 $\Delta^{*_\rh}=P^\rh_{0,1}(\Delta^*)=
       \rh^{-1}\circ\Delta^*\circ \rh$.

We shall now state the following lemma whose proof
we reserve for the appendix.

\begin{lemma}\label{lemma1}
Let $M$ be a manifold
provided with a volume form $\rh$. Denote by $\D(M)$ the
space of linear differential operators
on functions on $M$ and by $\D^{(n)}(M)$ its subspace of
differential operators of order $\leq n$.
  Let $F$ be the linear map defined by
         \begin{equation}\label{lemma1eq}
  F(\Delta)=a\Delta+b\Delta^{*_\rh}
+c\Delta(1)+d\Delta^{*_\rh}(1)
          \end{equation}
mapping $\D(M)$ to itself,
where $a,b,c,d$ are constants,
and $\Delta^{* \rh}$ is the operator
adjoint to $\Delta$ with respect to volume form $\rh$.
Then this map is $\SDiff_\rh(M)$-equivariant,
and moreover the converse implication is true
in the case that $M$ is connected manifold of
dimension $\geq 3$:
if $F$ is an arbitrary linear $\sdiff_\rh(M)$-equivariant
map defined on the subspace $\D^{(n)}(M)$
with values in the space $\D(M)$,
then $F(\Delta)$ has the appearance \eqref{lemma1eq}.
(We suppose that $F(\Delta)$ is a differential
polynomial on the coefficients of the operator $\Delta$.)

\end{lemma}

\begin{remark}
If our map $F$ obeys the stronger condition that it is equivariant with respect
to the whole algebra $\diff(M)$ of infinitesimal
diffeomorphisms then we have that
         \begin{equation}\label{simplelemma}
  F(\Delta)=a\Delta+c\Delta(1)\,.
          \end{equation}
\end{remark}

 The statement of the lemma and equation \eqref{simplelemma} are
a sort of Schur lemma for the
group of diffeomorphisms  and its subgroup $\Vol_\rh(M)$.
Equation
\eqref{simplelemma} states that the $\Diff(M)$-equivariant
linear maps on operators on functions
is proportional to the identity operator
 on the two invariant subspaces of normalised operators and operators
of multiplication by functions.
Equation
\eqref{lemma1eq} of the lemma states that a $\Vol_\rh(M)$-equivariant
linear map on $\D(M)$
is proportional to the identity operator
 on the invariant spaces of normalised self-adjoint
operators and normalised anti-self-adjoint
operators. (Operator $\Delta$
is normalised if
 $\Delta(1)=0$.)

We shall use the lemma for constructing  pencil liftings
which are $\vol(M)$ and $\diff(M)$-equivariant.
Firstly we consider $\vol(M)$-equivariant pencil liftings.

 Let a manifold $M$ be equipped with a volume form $\rh$.
 The  pencil lifting \eqref{canonicalpencillifting1}
 assigns to any operator $\Delta\in \D^{(n)}_\lo(M)$
 the operator $\widehat \Delta=\widehat {P^\rh_\lo}(\Delta)$
of the same order on algebra of densities $\F(M)$,
 i.e. the operator pencil
$\{\Delta_\l\}\colon \Delta_\l=P^\rh_{\l, \lo}(\Delta)$.
 This pencil passes through the operator $\Delta$,
 $\widehat\Delta\big\vert_{\hl=\lo}=\Delta$.
 Thus the canonical lifting \eqref{canonicalpencillifting1}
is $\vol_\rh(M)$-lifting
 which is strictly regular.
Are there another strictly regular or just regular
$\vol_\rh(M)$-liftings?  To answer this question we analyze regular
$\vol_\rh(M)$-lifting maps using lemma \ref{lemma1}.

 Pick arbitrary $n$ and $\l_0$ and let $\hPi=\hPi^{\rh,(n)}_\lo(\Delta)$
 be a regular $\vol_\rh(M)$-lifting
map defined on the space $\D_\lo^{(n)}(M)$.
Recall that a regular lifting on
$\D_\lo^{(n)}(M)$ is a linear map on $\D_\lo^{(n)}(M)$
which takes values in the space $\widehat{\D}^{(n)}(M)$
of operators of order $\leq n$
on $\F(M)$ and a strictly regular
lifting of $\D^{(n)}_\lo(M)$
restricted to the subspaces
$\D^{(k)}_\lo(M)$ takes values in the
subspaces $\widehat{\D^{(k)}}(M)$ for all $k\leq n$.

Using the identification isomorphisms \eqref{identification1}
 assign to the lifting  $\hPi^{\rh,(n)}_\lo$ the pencil $\{F^\hPi_\l\}$
of differential operators acting on functions, defined by
$F^\hPi_\l=P^\rh_{0,\l}\circ
   \left(\hPi^{\rh,(n)}_\lo\right)\big\vert_{\hl=\l}
   \circ P^\rh_{\lo,0}$, i.e.
for an arbitrary operator $\Delta$ on functions
               \begin{equation*}%\label{assigntoliftingfunction}
               F^\hPi_\l(\Delta)=
                  \rh^{-\l}\circ
   \hPi^{\rh,(n)}_\lo\big\vert_{\hl=\l}
\left(\rh^{\lo}\circ\Delta\circ\rh^{-\lo}\right)
                    \circ\rh^{\l}\,.
               \end{equation*}
All the maps $F^\hPi_\l$ are linear,
$\vol_\rh(M)$-equivariant maps since the
lifting $\hPi$ is a $\vol_\rh(M)$-lifting. Using equation
\eqref{lemma1eq} of lemma \ref{lemma1}
applied to the pencil $\{F^\hPi_\l\}$,
the relations \eqref{adjointforcanonical2},
and the fact  that the map $\hPi=\hPi^{\rh,(n)}_\lo$
is regular, we see that in the case if dimension of $M$ is greater than
2 then
        \begin{equation}\label{regularvollifting1}
       \hPi(\Delta)=A(\hl)\widehat {P^\rh_{\l_0}}
              \left(\Delta\right)+\,
            B(\hl)
            \left(\widehat {P^\rh_{\l_0}}\,
                 \left(\Delta\right)\right)^*+\,
                 C(\hl)\widehat {P^\rh_{\l_0}}(1)+
 D(\hl)\left(\widehat {P^\rh_{\l_0}}\,
                 \left(\Delta\right)\right)^*\!(1)\,.
                  \end{equation}
Here $A(\hl),B(\hl),C(\hl)$ and $D(\hl)$ are vertical
operators \eqref{verticaloperators}, that are
polynomials in $\hl$ of the form:
        \begin{equation*}\label{coefficientspolynomials}
A(\hl)=1-b(\hl-\lo)\,,\, B(\hl)=(-1)^nb(\hl-\lo)\,,\,
C(\hl)=\sum_{k=1}^n c_k (\hl-\lo)^k\,,\,
D(\hl)=\sum_{k=1}^n d_k (\hl-\lo)^k\,,
      \end{equation*}
where $\{b,c_k,d_k\}$, $k=1,2,\dots,n$ are constants.
Note that difference of two arbitrary lifting maps
\eqref{regularvollifting1} belong to linear space of liftings
which vanish the space $\D^{(n)}_\lo$.
We come to the following proposition

\begin{proposition}\label{proposition1}

1. All linear maps of the form \eqref{regularvollifting1}
are regular $\Vol_\rh(M)$-liftings
defined on the space $\D^{(n)}_\lo(M)$.
If $n\geq 2$, then the space of these liftings is an affine space of
dimension $2n+1$. Its dimension is equal to $2$ if $n=1$
(see the example \ref{example2} below for more detail).

2. If $M$ is a
connected manifold of dimension $\geq 3$ then
an arbitrary regular $\vol_\rh(M)$-equivariant lifting
of $\D^{(n)}_\lo$ belongs to this affine space

3. If in equation \eqref{regularvollifting1}
$b=0$, $d_1=-c_1$ and
$d_i=c_i=0$ for all $i\geq 2$ then we come to an
 affine line of strictly regular $\Vol_\rh(M)$-liftings.
 If $M$ is a connected manifold of dimension $\geq 3$ then
an arbitrary strictly regular
$\vol_\rh(M)$-equivariant liftings of $\D^{(n)}_\lo$
belongs to this affine line.

\end{proposition}

\begin{remark}\label{remarkonflag}
The affine space \eqref{regularvollifting1}
of $\vol_\rh(M)$-regular lifting maps
has the following natural flag structure:
\begin{itemize}

\item
point --- $A=1,B=C=D=0$, i.e. the
strictly regular lifting
$\Pi^{\rh,(n)}_\lo(\Delta)=P^\rh_\lo(\Delta)$,

\item
line ---  $C=D=0$, i.e. the 1-parameter family $\hPi_b$
of regular liftings
\begin{equation}\label{affinelineofliftings1}
\hPi_b(\Delta)=(1-b(\hl-\lo))\hPrh(\Delta)+
(-1)^nb(\hl-\lo)\left(\hPrh(\Delta)\right)^*\,,\,\,
 b\in\RR\,.
\end{equation}

\end{itemize}

\end{remark}

In the previous section we considered canonical liftings
for first order operators, in the following example
we will use the proposition
to describe all regular $\vol(M)$-equivariant
lifings for first order operators and distinguish those that are
$\diff(M)$-equivariant. As a consequence we will describe
all $\diff(M)$-liftings for first order operators
since any $\diff(M)$-lifting is necessarily a $\vol(M)$-lifting.

\begin{example}\label{example2}
 Choose an arbitrary volume form $\rh$ on $M$ and
pick an arbitrary $\lo$. Let $\Delta$ be
a first order
operator acting on densities of weight $\lo$,
$\Delta\in \D^{(1)}_\lo(M)$. According to equation \eqref{localexpression1}
for the canonical liftings $\hPrh$ \eqref{canonicalpencillifting1}
we have that
            $$
\hD=\hPrh(\Delta)=A^i(x)\left(\p_i+(\hl-\lo)\G_i(x)\right)+B(x)\,,
\,\, (\Delta=A^i(x)\p_i+B(x))\,,
           $$
where as usual $\Gamma_i(x)=-\p_i\log\rho(x)$ ($\rh=\rho(x)|Dx|$)
 are components of a flat
connection in the local coordinates $x^i$.
Hence using equation \eqref{regularvollifting1} we come
to the following family of regular $\vol_\rh(M)$-liftings
      $$
 \hPi(\Delta)=(1-b(\hl-\lo))\hPrh(\Delta)-
b(\hl-\lo)\left(\hPrh(\Delta)\right)^*+c(\hl-\lo)\hPrh(1)+
d(\hl-\lo)\left(\hPrh(\Delta)\right)^*\!(1)=
       $$
      \begin{equation}\label{volliftingsoffirstorderoperators1}
 A^i(x)\p_i+B(x)+(\hl-\l_0)
   \left(k_1B(x)+k_2\p_i A^i(x)+
 \left(1-\lo k_1-k_2\right)A^i(x)\Gamma_i(x)\right)\,, k_1,k_2\in \RR
     \end{equation}
  where we denote by $k_1=c+d-2b$ and $k_2=b-d$.
We come to a $2$-parametric family of regular $\vol_\rh(M)$-liftings, i.e an
affine plane of liftings.
It is evident that the above lifting
is strictly regular lifting if $k_1=0$.

Now we can calculate all regular
$\diff(M)$-lifting maps on $\D^{(1)}_\lo(M)$.
 We have that a regular $\diff(M)$-equivariant lifting is
 necessarily a $\vol(M)$-lifting that does not depend on the volume form.
 One can
see from equation \eqref{volliftingsoffirstorderoperators1}
that the  regular $\vol(M)$-liftings that do not depend on the
volume form if $k_2+\lo k_1-1=0$, i.e. $k_2=1-\lo k_1$.
  Thus we come to an affine line of $\diff(M)$-regular liftings:
      \begin{equation*}%\label{volliftingsoffirstorderoperators2}
\hPi(\Delta)= A^i\p_i+\hl\p_iA^i(x)+
    \left(1+k_1\left(\hl-\lo\right)\right)
   \left(B(x)-\lo\p_i A^i\right)\,, k_1\in \RR\,,
     \end{equation*}
which is just the family of  canonical liftings
\eqref{canonicalliftingoffirstorderoperator2}).
In the case that $k_1=0$ we come to the
strictly regular $\diff(M)$-lifting
(compare with the canonical liftings
\eqref{canonicalliftingoffirstorderoperator1}),
and finally in the case if $k_1=2/(2\lambda_{0}-1)$
we come to the anti-self-adjoint lifting
\eqref{canonicalliftingoffirstorderoperator3}
(for $\lo\not={1\over 2}$).

We see that that $\diff(M)$-equivariant liftings of
first order operators
are exhausted by the canonical liftings considered in the
section \ref{subsection2.1}.
Later we shall apply these methods to the analysis of liftings for higher order operators.

\end{example}

 \begin{example}\label{exampleofsecondorder1}
  Consider liftings of second order operators acting on
densities of weight $\lo$ and let
$\Delta=S^{ij}(x)\p_i\p_j+T^i(x)\p_i+R(x)\in\D^{(2)}_\lo(M)$.
Its adjoint is of the form
             $$
\Delta^*=S^{ij}(x)(x)\p_i\p_j+(2\p_rS^{ri}(x)-T^i)(x)\p_i+
 (R(x)-\p_rT^r(x)+\p_r\p_q S^{rq}(x))
             $$
acting on densities of weight $1-\lo$ (see \eqref{canonicalscalarproduct1}).
Consider the lifting  \eqref{canonicalpencillifting1}
of both these operators:
           $$
\widehat{P^\rh_\lo}(\Delta)=\rh^{\hl-\lo}\circ\Delta\circ
          \rh^{\lo-\hl}= S^{ij}(x)\p_i\p_j+2(\hl-\lo)\G^i(x)\p_i
  +T^i(x)\p_i+
                    $$
                 $$
                 +
    (\hl-\lo)^2\G^i(x)\G_i(x)+(\hl-\lo)\p_i\G^i(x)+
   (\hl-\lo)(T^i(x)-\p_rS^{ri}(x))\G_i(x)+R(x)\,,
           $$
      $$
\left(\widehat{P^\rh_\lo}\left(\Delta\right)\right)^*=
\left(\widehat{P^\rh_{1-\lo}}\left(\Delta^*\right)\right)=
   S^{ij}(x)\p_i\p_j+2(\hl+\lo-1)\G^i(x)\p_i
+(2\p_r S^{ri}(x)-T^i(x))\p_i+
                 $$
                 $$
    (\hl+\lo-1)^2\G^i(x)\G_i(x)+(\hl+\lo-1)\p_i\G^i(x)+
   (1-\hl-\lo)T^i(x)\G_i(x)+R(x)-\p_rT^r(x)+\p_r\p_qS^{rq}(x)\,.
           $$
Here $\G_i(x)=-\p_i\log \rho(x)$ and $\G^i(x)=S^{ik}(x)\G_k(x)$.

For the rest of this example, to ease the calculations,
we will suppose that local coordinate system $(x^i)$
are normal coordinates with respect to the
volume form $\rh$, i.e. $\rh$ is the
coordinate volume form in these coordinates
$\rh=|Dx|$. In particular in normal coordinates the
connection $\G_i(x)=-\p_i\log \rho(x)$ vanishes.
In this case the above formulae become
far simpler and we see that the regular $\Vol(M)$-lifting map
\eqref{regularvollifting1} has the appearance
        $$
\widehat\Pi(\Delta)=\Delta
+2b(\hl-\lo)\left(\p_rS^{ri}(x)-T^i(x)\right)\p_i+
 b(\hl-\lo)\left(\p_r\p_iS^{ri}(x)-\p_iT^i(x)\right)\p_i+
        $$
        $$
    (\hl-\lo)
                \left[
        \left(c_1+c_2(\hl-\lo)\right)R(x)+
        \left(d_1+d_2(\hl-\lo)\right)
        \left(\p_r\p_qS^{rq}(x)-\p_i T^i(x)+R(x)\right)
      \right]\,.
        $$
This regular lifting depends on $5$ parameters and
in general it is not strictly regular.
For example take the first order operator
$\Delta=T^i(x)\p_i+R(x)\in\D^{(2)}_\lo(M)$ ($S^{ij}(x)\equiv 0$),
then the lift is of the form
             $
\widehat\Pi(\Delta)=\Delta-2b(\hl-\lo)T^i(x)\p_i+\dots
             $,
 which is an operator of order $2$ (if $b\not=0$
and $T^i(x)\not=0$).
We find that this lifting is strictly regular
if $b=c_2=d_2=0$ and $c_1=-d_1$:
        $
\Pi(\Delta)=\Delta+
d(\hl-\lo)\left(\p_iT^i-\p_r\p_qS^{rq}\right)$.
 \end{example}

\section{Self-adjoint $\vol_\rh(M)$-liftings
and $\diff(M)$-liftings}\label{section4}

To use the results obtained above it is useful to
think of $\diff(M)$-liftings as $\vol_\rh(M)$-liftings
which are independent
of the volume form $\rh$. We will use this
fact to describe $\diff(M)$-liftings
for higher order operators in this section.
(For first order operators we did this
in example \ref{example2}.)

First we consider $\vol(M)$-liftings,
 which are self-adjoint or anti-self-adjoint, depending
on whether the number $n$ is even or odd (see remark
 \ref{why operator is self-adjoint?}).
The condition of (anti)-self-adjointness
arises naturally when one tries to minimise
the dependance of lifting maps on a volume form.

We show that if a lifting is invariant under a variation of the volume form
(up to a vertical map) then it is (anti)-self-adjoint.
This will lead us to give an alternative proof of existence and uniqueness
 of $\diff(M)$-lifting for second order operators
(this is just the self-adjoint lifting
described in subsection \ref{subsection2.2})
 and we will show that there are no
$\diff(M)$-liftings of operators of order $\geq 3$ for
manifolds of dimension $\geq 3$.

\subsection{Self-adjoint and anti-self-adjoint $\vol(M)$-liftings}

As usual choose an arbitrary volume form $\rh$ on manifold $M$.
Pick an arbitrary $n$ and $\lo$.
In the flag of regular $\sdiff_\rh$-liftings
of space $\D^{(n)}_\lo(M)$
consider the line of liftings \eqref{affinelineofliftings1},
 $\hPi_b=\hPi^{\rh,(n)}_{\lo,b}(\Delta), b\in \RR$
(see remark \ref{remarkonflag}).
In the case that $n$ is an even number,
choose, in this family,
the self-adjoint
lifting of the space $\D^{(n)}_\lo$
(in the case where $n$ is odd we choose the anti-self-adjoint lifting).
 The condition that $\left(\hPi_b(\Delta)\right)^*=
(-1)^n\hPi_b(\Delta)$ implies that
$(1-b(\hl-\lo))^*=b(\hl-\l_0)$ i.e.
   \begin{equation}\label{necessarycondition1}
          b={1\over 1-2\lo}\,.
      \end{equation}
Hence in the case that $\lo\not={1\over 2}$,
we come to a distinguished pencil lifting map of $\D^{(n)}_\lo$
belonging to the affine line
\eqref{affinelineofliftings1}:
                \begin{equation}\label{distinguishedlifting}
 \hPi_{\rm disting.}(\Delta)=
 {\hl+\lo-1\over 2\lo-1}\hPrh\left(\Delta\right)+
 (-1)^n{\lo-\hl\over 2\lo-1}
 \left(\hPrh\left(\Delta\right)\right)^*\,,
              \end{equation}
which is (anti)-self-adjoint:
$\left(\hPi_{\rm disting.}(\Delta)\right)^*=
 (-1)^n\hPi_{\rm disting.}(\Delta)$.

\begin{example} Take a $3$-rd order operator
         $$
      \Delta=S^{ikm}\p_i\p_k\p_m+G^{ik}\p_i\p_k+A^i\p_i+R\,,
              $$
acting on densities of weight $\lo$,
$\Delta\in \D^{(n)}_\lo$ where $\lo\not={1\over 2}$.
We choose normal cooridnates, i.e. coordinates such that volume form
$\rh=|Dx|$. Then
      $$
               \Pi_{\rm disting.}(\Delta)=
  {\hl+\lo-1\over 2\lo-1}\hPrh\left(\Delta\right)-
 {\lo-\hl\over 2\lo-1}
 \left(\hPrh\left(\Delta\right)\right)^*=
       $$
               \begin{equation*}
            {\hl+\lo-1\over 2\lo-1}
  \left(S^{ikm}\p_i\p_i\p_m+
 G^{ik}\p_i\p_k+\dots\right)+
        {\hl-\lo\over 2\lo-1}
 \left(-S^{ikm}\p_i\p_k\p_m+
 \left( G^{km}-3\p_{i}S^{ikm}\right)\p_k\p_m+
          \dots\right)=
            \end{equation*}
               \begin{equation*}
 S^{ikm}\p_i\p_k\p_m+
3{\hl-\lo\over 2\lo-1}\p_{i}S^{ikm}\p_k\p_m+
 {2\hl-1\over 2\lo-1}G^{km}\p_k\p_m+\dots
               \end{equation*}
(we denote by dots operators of order $\leq 1$).
\end{example}

 Now consider an arbitrary (anti)-self-adjoint liftings
 in the affine space \eqref{regularvollifting1},
which differ from the distinguished lifting
\eqref{distinguishedlifting}
by a vertical map $C(\hl)\hPrh(\Delta)(1)+
D(\hl)\left(\hPrh(\Delta)\right)^*(1)$ (see equation
\eqref{regularvollifting1}).
 The polynomials $C(\hl)$ and $D(\hl)$
 in equation \eqref{regularvollifting1}
have to be self-adjoint vertical operators
 if $n$ is even, respectively anti-self-adjoint if $n$ is odd.
 One can see that these polynomials have the following appearance:
         \begin{equation}\label{scalartermspolynomials1}
         C(\hl)=
             t^{p(n)}\sum_{k=1}^{n-p(n)\over 2}
         c_k \left(t^{2k}(\hl)-t^{2k}(\lo)\right)\,,\,\,
D(\hl)=t^{p(n)}\sum_{k=1}^{n-p(n)\over 2}
         d_k \left(t^{2k}(\hl)-t^{2k}(\lo)\right)\,,
         \end{equation}
where $t(\hl)=\hl-{1\over 2}$ is anti-self-adjoint
linear polynomial in $\hl$:
 $t^*(\hl)=\left(\hl-{1\over 2}\right)^*
=-\hl+{1\over 2}=-t(\hl)$,
$p(n)=0$ if $n$ is even and $p(n)=1$ if $n$ is odd.

\begin{proposition}\label{proposition2}

For ${\lo\not={1\over 2}}$ and $n\geq 2$ the $2n+1$-dimensional
affine space of
regular $\vol_\rh(M)$-liftings of the space $\D^{(n)}_\lo$
\eqref{regularvollifting1}
possesses a subplane of self-adjoint
(anti-self-adjoint) liftings of
dimension $n$ (of dimension $n-1$)
if $n$ is even (if $n$ is odd) described
by the equations \eqref{scalartermspolynomials1}.

The affine line \eqref{affinelineofliftings1}
of $\vol(M)$-liftings  possesses
a distinguished
lifting $\hPi_{\rm disting.}(\Delta)$ \eqref{distinguishedlifting}.
This lifting is self-adjoint if $n$ is even, and it is anti-self-adjoint
if $n$ is odd.
\end{proposition}

\begin{example}\label{exampleofsecondorder2} In
example \ref{exampleofsecondorder1}
we considered regular liftings for second order operators. We now consider
self-adjoint regular liftings of these operators.
For an arbitrary second order
operator acting on densities of weight $\l_0$,
$\Delta=S^{ij}(x)\p_i\p_j+T^i(x)\p_i+R(x)\in\D^{(2)}_\lo$
($\lo\not=1/2$)
according to the proposition
self-adjoint lifting $\hPi$ has the following appearance:
               $$
\hPi(\Delta)=\hPi_{\rm disting.}(\Delta)+
\left(t^2(\hl)-t^2(\lo)\right)
\left(cR+d\left(\p_r\p_i S^{ri}-\p_iT^i+R\right)\right)\,.
               $$
Using the equations for regular liftings from
example \ref{exampleofsecondorder1} and equation
\eqref{distinguishedlifting},
we have that
         $$
\hPi_{\rm disting.}(\Delta)=
 {\hl+\lo-1\over 2\lo-1}\hPrh\left(\Delta\right)+
 {\lo-\hl\over 2\lo-1}
 \left(\hPrh\left(\Delta\right)\right)^*
         $$
 and
         $$
          \hPi(\Delta)=
   S^{ij}(x)\p_i\p_j+\p_rS^{ri}\p_i+
   {2\hl-1\over 2\lo-1}(T^i-\p_rS^{ri})
      +{\hl-\lo\over 2\lo-1}(\p_iT^i-\p_r\p_i S^{ri})+R+
                $$
                \begin{equation}\label{forthenextexample1}
                +
   \left(\hl\left(\hl-1\right)-
     \lo\left(\lo-1\right)\right)
     \left(cR+d\left(\p_r\p_i S^{ri}-\p_iT^i+R\right)\right)\,,
         \end{equation}
where $c,d$ are arbitrary constants
(We work in normal coordinates such that volume form $\rh=|Dx|$.).
We come to the plane ($2$-dimensional space) of self-adjoint
regular $\sdiff_\rh$ lifting maps of $\D^{(2)}_\lo(M)$ for $\lo\not=1/2$.
Note that in the case if $\lo\not=0,1$
this lifting map
can be presented as the sum of canonical map
\eqref{selfadjointoperator0} ($F=0$) and vertical map
             $$
\hPi(\Delta)=\left(\hl\left(\hl-1\right)-\lo\left(\lo-1\right)\right)
\left(\left(c+d)\lo(\lo-1)-1\right)\theta
 \left(c-d)\lo+d\right)\p_i\gamma^i\right)\,,
         $$
where
$\gamma^i$, $\theta$ are upper connection and Branse-Dicke scalar
corresponding to operator $\Delta$
(see section \ref{subsection2.2}).

\end{example}

\subsection {Dependance of $\vol(M)$-liftings
and (anti)-self-adjoint liftings on the volume form}

Now we shall calculate the infinitesimal variation
of regular $\vol(M)_\rh$-equivariant
liftings \eqref{regularvollifting1} with respect to an infinitesimal
variation of the volume form.

Pick an arbitrary $n$ and $\lo$ and choose
within the affine space of liftings
\eqref{regularvollifting1} an arbitrary
regular $\vol_\rh(M)$-lifting, $\hPi$.
  We shall calculate the variation
of this lifting with respect to the
variation of the volume form, $\rh\to \rh+\delta\rh$,
firstly by considering the canonical
$\vol_\rh(M)$-lifting $\widehat P^\rh_\lo$,
(see \eqref{canonicalpencillifting1})
which has the following variation:
        \begin{equation}\label{infinitesimalchangingofcanonicalpencil1}
 \delta_\rh \hPrh=
  \left(\hl-\lo\right)
       \left(
\rh^{-1}\delta\rh\circ \hPrh
               -
\hPrh\circ\delta\rh^{-1}\rh
       \right)
        =
   \left(\hl-\lo\right)
     \left[\rh^{-1}\delta\rh,\hPrh\right]\,.
           \end{equation}
(If $\rh=\rh_t$ is a one-parametric family of volume forms,
then $\dot \rh_t=
\rh_t\left(\rh_t^{-1}\dot\rh_t\right)$, where $\rh_t^{-1}\dot\rh_t$
is a scalar function and $\delta\rh=
\dot\rh_t\big\vert_{t=0}\delta t$.)

Any lifting \eqref{regularvollifting1}
is the sum of a lifting $\hPi_b$
belonging to the affine line \eqref{affinelineofliftings1},
and a vertical map (see Proposition\ref{proposition1}).
The variation of a vertical map with respect to volume form
is a vertical map,
 hence we have that the variation
of the regular $\sdiff_\rh$-lifting \eqref{regularvollifting1}
with respect to volume form $\rh$ is equal to
      $$
\delta_\rh \Pi\colon\quad
\delta_\rh\Pi(\Delta)=
\delta_\rh\left(\hPi_b(\Delta)\right)
+\hbox{variation of vertical map}=
     $$
     $$
A(\hl)\left(\hl-\lo\right)
  \left[\rh^{-1}\delta\rh, \hPrh(\Delta)\right]+
B(\hl)\left(\hl+\lo-1\right)
  \left[\rh^{-1}\delta\rh, \hPrh^*(\Delta)\right]
+\hbox{vertical map}
      $$
          \begin{equation*}
 =
\left(1+b\left(2\lo-1\right)\right)
 \left(\hl-\lo\right)
      \left[\rh^{-1}\delta\rh, \hPrh\left(\Delta\right)\right]-
\end{equation*}
\begin{equation}\label{variationofregularlifting3}
          +
      b(1-\hl-\lo) \left(\hl-\lo\right)
\underbrace {
\left[\rh^{-1}\delta\rh\,,P^\rh_\lo\left(\Delta\right)
  -(-1)^n\left(P^\rh_\lo\right)^*\left(\Delta\right)\right]+
            }_{\hbox{operator of order $\leq n-2$}}
           \hbox{vertical map}
\end{equation}
(Recall that $(\hl-\lo)^*=(1-\hl-\lo)$, the operator
$P^\rh_\lo(\Delta)-(-1)^n\left(P^\rh_\lo(\Delta)\right)^*$ has order
$\leq n-1$, and the commutator of two operators of orders
$m,n$ has order $m+n-1$).)

It follows from this relation that the condition
$b={1\over 1-2\lo}$ (if $\lo\not=1/2$)
 \eqref{necessarycondition1}, which means
that the lifting is the distinguished lifting up to a vertical map,
is a necessary condition for the lifting to be independent of infinitesimal
variations of the volume form
(up to a vertical maps).
We come to the following proposition:

  \begin{proposition}\label{proposition3}

  Choose a volume form $\rh$ on manifold $M$.

1. Let $\Pi=\Pi_{\rm disting.}$
be the distinguished $\sdiff_\rh(M)$-lifting
\eqref{distinguishedlifting} defined on the space $\D^{(n)}_\lo(M)$
($\lo\not={1/2}$).

 Then
         \begin{equation}\label{distinguished4}
\delta_\rh\Pi(\Delta)={(\hl-\lo)(\hl+\lo-1)\over 2\lo-1}
   \left[\rh^{-1}\delta\rh,\hPrh(\Delta)-(-1)^n
       \left(\hPrh(\Delta)^*\right) \right]\,.
    \end{equation}

2.Let $\hPi$ be an arbitrary
regular $\vol_\rh(M)$-equivariant
lifting defined on the space $\D^{(n)}_\lo(M)$.
Suppose that $n\geq 2$, and let $\Delta_0$ in
$\D^{(n)}_\lo(M)$
be an operator of order $n$, i.e.
$\Delta_{0} \not \in \D^{(n-1)}_\lo(M)$,
such that the infinitesimal variation
with respect to the volume form $\rh$
of $\Pi$ at $\Delta_0$ vanishes:
         $$
\delta_\rh\Pi\big\vert_{\Delta=\Delta_0}=0\,.
 $$
Then in the case the dimension of $M$ is greater than 2

\begin{itemize}

\item

$\lo\not={1\over 2}$ since in the case of $\lo={1\over 2}$
the $\sdiff$-lifting of $n$-th order operator essentially depends
on the volume form.

\item

the lifting $\Pi$ is equal to the distinguished lifting
\eqref{distinguishedlifting} up to vertical maps:
       \begin{equation*}\label{distinguishedlifting2}
\hPi(\Delta)=
 \hPi_{\rm disting.}(\Delta)+\hbox{vertical map}\,,
\quad
\hPi^*(\Delta)=(-1)^n \hPi(\Delta)+\hbox{vertical map}\,.
 \end{equation*}
In particular  the canonical lifting of the operator
$\Delta_0$  is (anti)-self-adjoint up to a vertical operator.

\end{itemize}

\end{proposition}

  Namely if an operator $\Delta_0$ has order $n$,
($\Delta_0\in \D^{(n)}_\lo(M)$, and
 $\Delta_0\not\in \D^{(n-1)}_\lo(M)$)
then due to equation \eqref{variationofregularlifting3}
the infinitesimal variation of the canonical lifting $P^\rh_\lo$
at $\Delta_0$ does not vanish,
$[\delta_\rh,P^\rh_\lo(\Delta_0)]\not=0$, if
$(1+b(2\lo-1))\not=0$ i.e. if condition \eqref{necessarycondition1}
is not obeyed, and the pencil is not the distinguished pencil.
In particular if $\lo=1/2$ then
the variation does not vanish for any choice of $b$.
If $\lo\not=1/2$ then the variation has a chance to vanish
only for the distinguished pencil.

This proposition is crucial for extracting
$\diff(M)$-equivariant liftings from the class of
$\vol(M)$-liftings.
To this end let $\Pi^{\rh}$ be an arbitrary $\vol_\rh$-lifting
\eqref{regularvollifting1} and let $\X$ be an arbitrary vector field on $M$.
We can express the Lie derivative $\widehat {\ad_X}$ of the lifting
 $\Pi^{\rh}$ in terms of the variation with respect to the volume form.
The Lie derivative, $\ad_X$, of an arbitrary operator
is equal to
   $\ad_X\Delta\colon\quad
      \ad_\X\Delta(\bs)=
 \L_\X\left(\Delta\left(\bs\right)\right)-
  \left(\Delta\left(\L_\X\bs\right)\right)$
          and for maps on operators
\begin{equation*}
   \label{liederivativeofmaponoperators}
\widehat{\ad_\X}\hPi\colon\,\,
\widehat{\ad_\X}\hPi \left(\Delta\right)=
\ad_\X\left(\hPi\left(\Delta\right)\right)-
 \left(\hPi\left(\ad_\X\Delta\right)\right)\,.
\end{equation*}

If $\hPi=\hPi^\rh$ is a map of operators
depending on the volume form $\rh$,
such that for an arbitrary volume form it is
$\vol_\rh(M)$-equivariant, then for a vector field $\X$
we arrive at the Lie derivative from the
variation by setting $\delta\rh$ equal to
$\L_\X\rh=\rh{\rm div}_\rh \X$:
   \begin{equation}\label{isittrueingeneral?}
   \widehat {\ad_\X}\hPi^\rh=
\delta_\rh\hPi^\rh\big\vert_{\delta\rh\mapsto
          \rh{\rm div}_\rh\X}\,.
   \end{equation}
where ${\rm div}_\rh\X$ is the divergence of $\X$ with
respect to $\rh$, see equation \eqref{divergenceless}:
If $\rh=\rh(t)$ is one-parametric family
then $\rh{\rm div}_\rh\X=\dot\rh_t\big\vert_{t=0}$ and
$\delta\rho=\dot\rh_t\big\vert_{t=0}\delta t$.
  This formula can be easily checked for the canonical lifting
 $\hPrh$ and hence for an arbitrary lifting \eqref{regularvollifting1}
 in  the affine space of regular $\vol(M)$-equivariant liftings:
          $$
   \widehat{\ad_\X}\hPrh(\Delta)=
\ad_\X\left(\rh^{\hl-\lo}\circ\Delta\circ\rh^{\lo-\hl}\right)
         -
\rh^{\hl-\lo}\circ\ad_\X\Delta\circ\rh^{\lo-\hl}=
       \left(\hl-\lo\right)\left[{\rm div\,}_\rh\X,
        \hPrh(\Delta)\right]\,.
          $$
(Compare with equation
\eqref{infinitesimalchangingofcanonicalpencil1}.)
\begin{corollary}\label{corollary1}
   Let $\hPi=\hPi^{\rh,(n)}_\lo$ be a regular
$\sdiff_\rh(M)$-lifting
defined on $\D^{(n)}_\lo(M)$. Let the map $\hPi$ be
 $\diff(M)$-equivariant
at the given operator $\Delta_0$ of order $n$:
  $$
\forall \X\,,
\,\widehat{\ad_\X}\hPi (\Delta)\big\vert_{\Delta=\Delta_0}=0\,.
 $$
Then in the case the dimension of the manifold $M$ is greater than 2,
the weight
$\lo\not={1\over 2}$ and the lifting
$\hPi$ is the distinguished (anti)-self-adjoint  map
up to a vertical map
(see equation \eqref{distinguishedlifting}). In particular
                     $$
            \hD_0^*=(-1)^n\hD_0+{\rm vertical\,\,operators}.
                    $$
\end{corollary}

We use this corollary to study $\diff(M)$-liftings for
higher order operators.
(Regular $\diff(M)$-invariant pencil liftings
for operators of order $1$ were already described
in  section \ref{subsection2.1} and example \ref{example2}).

\subsection
{ Regular $\diff(M)$-liftings for second order operators.}
\label{subsection4.3}

First we show that for weights $\l_0\not=0,1/2,1$
on manifolds of dimension $\geq 3$
the self-adjoint canonical pencil lifting
described in section \ref{subsection2.2}
is unique, regular and $\diff(M)$-equivariant lifting map
on operators of order $\leq 2$.
   (Note that this lifting's
 uniqueness can be obtained using the isomorphisms
   obtained by Duval and Ovsienko \cite{DuvOvs},
   and our analysis  gives a
complementary geometric picture behind their results.)
Also we will explain why there is no $\diff(M)$-equivariant lifting for
the exceptional weights $0,1/2$ and $1$.

 Recall that for $\l_0\not=0,1/2,1$
one can consider the canonical self-adjoint lifting
map on $\D^{(2)}_\lo(M)$ \eqref{selfadjointoperator0},
which sends the operator
$\Delta=S^{ij}(x)\p_i\p_j+T^i(x)\p_i+R(x)\in \D^{(2)}_\lo(M)$ to the pencil
            \begin{equation}\label{selfadjointoperator2}
  \hPi_{\rm can}(\Delta)=S^{ij}\p_i\p_j+\p_jS^{ij}\p_i+
       (2\hl-1)\gamma^i\p_i+\hl\p_i\gamma^i+\hl(\hl-1)\theta\,,
       \end{equation}
where the upper connection $\gamma^i$ and the Branse-Dicke function
$\theta$ are equal to
    \begin{equation}\label{geometricaldata4}
\gamma^i={T^i-\p_jS^{ij}\over 2\lo-1}, \,
     \theta={1\over \lo(\lo-1)}
   \left(
 R-{\lo\left(\p_iT^i-\p_ip_jS^{ij}\right)
      \over 2\lo-1}\right)\,.
 \end{equation}
The map \eqref{selfadjointoperator2} defines a regular
$\diff(M)$-equivariant lifting, and moreover this lifting is self-adjoint.
(See equations \eqref{selfadjointoperator0},
\eqref{condition2} and \eqref{normalisationcondition2}
in subsection\ref{subsection2.2}).

 Let $\hPi$ be an arbitrary regular $\diff(M)$-equivariant lifting
defined on $\D^{(2)}_\lo(M)$.
We then already know that
$\lo\not={1\over 2}$ (see corollary \ref{corollary1})
and we will show that the lifting,
$\hPi$, coincides with the canonical lifting
$\hPi_{\rm can.}$ (if $\dim M\geq 3$.)

Choose an arbitrary volume form
$\rh$. The lifting $\hPi$ has to be a regular $\vol_\rh(M)$-lifting,
since it is a $\diff(M)$-lifting.
Hence it follows from Corollary \ref{corollary1} that
       \begin{equation}\label{liftingforsecondorderoperator4}
\hPi(\Delta)=\underbrace
             {
   {\hl+\lo-1\over 2\lo-1}\hPrh(\Delta)+
   {\lo-\hl\over 2\lo-1}\left(\hPrh\left(\Delta\right)\right)^*
             }_
    {\hbox {distinguished lifting
$\hPi^{\rh,(2)}_{\lo,{\rm disting.}}(\Delta)$}}
             +
          \underbrace
            {
      C(\hl)\hPrh(\Delta)(1)+
 D(\hl)\left(\hPrh\left(\Delta\right)\right)^*(1)
            }_
           {\hbox{vertical map} }\,.
      \end{equation}
Using calculations in example \ref{exampleofsecondorder1}
collect the terms in the right hand side of this expansion,
which are proportional to $\p_i\Gamma^i(x)$
and to $\Gamma^i(x)\Gamma_i(x)$,
where $\Gamma_i(x)=-\p_i\log\rho(x)$ is connection of volume form,
$\rh$ ($\Gamma^i(x)=S^{ik}(x)\Gamma_k(x)$). We come to equations
          $$
       \begin{cases}
-\l_0 C(\hl)+(\l_0-1)D(\hl)=0\,,\quad
    \hbox{(terms proportional to $\p_i\Gamma^i(x)$)}\cr
\l_0^2 C(\hl)+(\l_0-1)^2D(\hl)=(\hl-\l_0)(\hl+\lo-1)\,,
    \hbox{(terms proportional to $\Gamma^i(x)\Gamma_i(x)$)}\cr
         \end{cases}
          $$
These condtions uniquely define polynomials $C(\hl)$ and $D(\hl)$.
Thus independence on volume form implies that only
one $\sdiff_\rh$-map $\hPi(\Delta)$ may be $\diff(M)$-equivariant.
This is the canonical map $\hPi_{\rm can.}$.
Hence $\hPi=\hPi_{\rm can.}$
\begin{remark}
We wish to calculate explicitly the vertical map in
\eqref{liftingforsecondorderoperator4}.
It is uniquely defined since it is equal to the
difference between the canonical map
\eqref{selfadjointoperator2} and the distinguished map.
If $\Gamma_i=-\log\rho(x)$ is the connection associated to $\rh$,
then using \eqref{localexpression1} one
can see that
      $$
    \underbrace
            {
      C(\hl)\hPrh(\Delta)(1)+
 D(\hl)\left(\hPrh\left(\Delta\right)\right)^*(1)
            }_
           {\hbox{vertical map} }=
\hPi_{\rm can.}(\Delta)-\hPi^{\rh}_{\lo,{\rm distinguish.}}(\Delta)=
          $$
       $$
     (\hl(\hl-1)-\lo(\lo-1))\left(\theta-2\Gamma_i\gamma^i+
           S^{ij}\Gamma_i\Gamma_j\right)\,.
      $$
 We come to the cocycle like object
$c_\rh(\Delta)=
\theta(x)-2\gamma^i(x)\Gamma_i(x)+S^{ij}(x)\Gamma_i(x)\Gamma_i(x)$.

\end{remark}
Now we consider the exceptional weights.
 First consider $\lo=1/2$.
 For second order operators acting on half-densities
their adjoint acts on half-densities too.
 If $\Delta=\Delta^*$ is a self-adjoint operator then
 choosing an arbitrary upper connection $\gamma^i$
 one can consider different self-adjoint operator
 pencils passing through this operator $\Delta$ (see in more details in \cite{KhVor4}).
  On the other hand there is no regular
$\diff(M)$-lifting defined on operators
 acting on half-densities, not even on the subspace of
self-adjoint operators. Indeed every $\diff(M)$-lifting
 has to be a $\sdiff_\rh(M)$-lifting, but
 in this case the lifting
 essentially depends on the volume form
(see Proposition \ref{proposition3}).

 Now consider the case $\l_0=0$.
 An arbitrary second order operator acting on
functions $\Delta=S^{ij}\p_i\p_j+T^i\p_i+R$
 is defined uniquely by a contravariant symmetric
tensor field $S^{ij}$, an upper connection
 $\gamma^i=\p_rS^{ri}\p_i-T^i$ and a scalar function $R=\Delta(1)$.
 An operator $\Delta$ is the sum of a normalised operator
 $\Delta_{\rm norm}=S^{ij}\p_i\p_j+T^i\p_i$, ($\Delta_{\rm norm}(1)=0$)
 and a scalar function
 $R(x)=\Delta(1)$. The
 $\diff(M)$-module $\D^{(2)}(M)$ of second order operators
 is the direct sum of the module
 of normalised operators and
 scalar functions.

  If $\theta$ is an
 arbitrary Branse-Dicke function corresponding to
an upper connection $\gamma^i$
 (see Appendix \ref{appendix1})
then according to \eqref{selfadjointoperator0}
 one can consider the self-adjoint canonical operator pencil
        \begin{equation*}\label{selfadjointontfucntions1}
        \Pi_{0,{\rm can}}(\Delta)=
S^{ij}\p_i\p_j+\p_rS^{ri}\p_i+(2\hl-1)\gamma^i\p_i+\hl\p_i\gamma^i+
        \hl(\hl-1)\theta
        \end{equation*}
        passing through the normalised operator $S^{ij}\p_i\p_j+(\p_rS^{ri}\p_i-\gamma^i)\p_i$.
This formula does not define a linear
map even on the submodule of normalised operators
($\Pi_{0,{\rm can}}(\Delta_1+\Delta_2)\not=
\Pi_{0,{\rm can}}(\Delta_1)+\Pi_{0,{\rm can}}(\Delta_2)$).

Choose an arbitrary volume form
$\rh$ and consider an arbitrary regular $\vol_\rh(M)$-lifting
on $\D^{(2)}(M)$.
We set $\l_0=0$ in \eqref{liftingforsecondorderoperator4}
and come to the linear map
                      $$
       \hPi^\rh(\Delta)=(1-\hl)\hPrh(\Delta)+
\hl\left(\hPrh\left(\Delta\right)\right)^*+
       C(\hl)\hPrh(\Delta)(1)+D(\hl)
\left(\hPrh\left(\Delta\right)\right)^*\!(1)\,.
                      $$
Now let us see how this lifting depends on $\rh$.
Using equation \eqref{localexpression1}
for the canonical lifting \eqref{canonicalpencillifting1}
   we have
    \begin{equation}\label{liftingofoepratorsonfunctions1}
       \begin{matrix}
\hPi^\rh(\Delta)=S^{ij}\p_i\p_j+\p_rS^{ri}\p_i+
(2\hl-1)\gamma^i\p_i+
   \hl\p_i\gamma^i+\cr
       +\hl(\hl-1)\left(2\gamma^i\Gamma_i-
\Gamma^i\Gamma_i\right)+
        (1+C(\hl)+D(\hl))R(x)+D(\hl)
(\p_i\gamma^i-\p_i\Gamma^i-
\gamma^i\Gamma_i+\Gamma^i\Gamma_i)\,,\cr
      \end{matrix}
              \end{equation}
where as usual $\Gamma_i(x)=-\p_i\log\rho(x)$ is the flat connection
induced by  the volume form.
One can see from this equation that the map $\hPi^\rh$
depends on the choice of a volume form
via the flat connection $\Gamma_i$
for arbitrary vertical operators $C(\hl)$ and $D(\hl)$.
(Terms proportional to $\Gamma^i(x)\Gamma_i(x)$
vanish only if $D(\hl)\not=0$, and the terms proportional
to $\p_i\Gamma^i(x)$, $\p_i\gamma^i(x)$
vanish if $D(\hl)\equiv 0$.)
 This dependance still exists if we define the
map only on normalised operators.
This implies that there are no $\diff(M)$-lifting maps on
$\D^{(2)}_0$, nor its subspace of normalised operators.

Analogous arguments work on the space of operators
on densities of weight $\lo=1$, since these operators are conjugate
to operators on functions.

The following
"limit" construction
may, however, be of interest
\begin{example}
 For arbitrary small $\lo$  consider the composition of
 the canonical volume form lifting $P^\rh_{\lo,0}$.
We come to
            $$
            \begin{matrix}
\hPi_{\rm can}\circ P^\rh_{_{\lo,0}}
\left(S^{ij}(x)\p_i\p_j+T^i(x)\p_i+R(x)\right)=\cr
=S^{ij}(x)\p_i\p_j+\p_rS^{ri}(x)\p_i+(2\hl-1)\gamma^i(x)+
  \hl\p_i\gamma^i(x)+\hl(\hl-1)\theta(x)\,,\cr
 \end{matrix}
            $$
where
         $$
         \begin{cases}
        \gamma^i=\gamma^i(\lo)=\p_rS^{ri}(x)-T^i(x)+O(\lo)\,,
          \cr
     \theta=\theta(\lo)=
-{R(x)\over \lo}+R(x)+\p_i\gamma^i(x)-\p_i\Gamma^i(x)+
    \gamma^i(x)\Gamma_i(x)+O(\lo)\,,\cr
         \end{cases}
          $$
Taking the limit $\lo\to 0$ we come
to a self-adjoint $\vol_\rh(M)$-lifting
defined on the subspace of normalised operators ($R=0$):
                 $$
 \hPi(\Delta)=\lim_{\lo\to 0}\hPi_{\rm can}\circ P^\rh_{_{\lo,0}}
\left(\Delta\right)=
S^{ij(x)}\p_i\p_j+\p_rS^{ri(x)}\p_i+(2\hl-1)\gamma^i(x)+
\hl\p_i\gamma^i(x)
    +\hl(\hl-1)\theta_\rh(x)\,,
                 $$
with $\gamma^i(x)=\p_pS^{pi}(x)-T^i(x)$ and
 $\theta_\rh=\p_i\gamma^i(x)-\p_i\Gamma^i(x)+
    \gamma^i(x)\Gamma_i(x)$.
($\theta_\rh(x)$ is
Branse-Dicke function corresponding to the upper connection
$\gamma^i$
(see Appendix \ref{appendix1}:
$\theta_\rh-2\gamma^i\Gamma_i+
\Gamma^i\Gamma_i={\rm div\,}_\rh (\gamma-\Gamma)$).)
This exceptional $\vol_\rh(M)$-lifting is self-adjoint.

\end{example}

\subsection{Regular $\diff(M)$-liftings for operators of order $\geq 3$}

\begin{corollary}\label{corollary2}
If $M$ is a manifold of dimension $\geq 3$, then
for an arbitrary weight $\lo$,
regular $\diff(M)$-equivariant liftings defined on
$\D^{(n)}_\lo(M)$ do not exist if $n\geq 3$.
 \end{corollary}

\proof  Suppose that $\hPi$ is a $\diff(M)$-equivariant lifting
defined on $\D^{(n)}_\lo(M)$.
  Choose an arbitrary volume form on $M$ then $\Pi$ has to be
$\vol_\rho(M)$-equivariant lifting. Due to the proposition
\ref{proposition3} and corollary\ref{corollary1}
the lifting $\Pi$ is, up to a vertical map,
the distinguished lifting $\Pi_{\rm disting.}$.
Consider the variation \eqref{distinguished4} of this map with
respect to a variation of the volume form.
This condition cannot be satisfied for an arbitrary operator
of order $n-2$. Hence we come to contradiction.

This result is known from the works \cite{LecomteMath1}
and \cite{Math2}. However this does not
exclude the existence of $\diff$-lifting maps
on some subspaces of differential operators
of an arbitrary order $n$. Based on the results of corollary \ref{corollary1}
consider the following example
\begin{example}
Consider $M=\RR^d$, $d$-dimensional affine space.
(We suppose that all functions and densities are rapidly
decreasing at inifinity)
Choose a volume form $\rh$ such that in a given
 Cartesian coordinates it is the coordinate volume form $\rh=|Dx|$.
(In arbitrary Cartesian coordinates $x^{i'}$ it will be equal to
 $\rh=C|Dx'|$, where $C$ is a constant).

  Operators in $\RR^d$ can be identified with contravariant symmetric
tensors.
Consider the subspace of symmetric divergenceless
contravariant symmetric tensors:
       \begin{equation}\label{divergenceless22}
\p_{i_k}S^{i_1\dots i_k}(x)\equiv 0\,.
\end{equation}
Pick an arbitrary $\l_0\not=1/2$.
 We denote by $L_+$ the linear space of operators formed by symmetric
contravariant divergenceless tensors of even rank,
and
by $L_-$ the linear space of
operators formed by symmetric
contravariant divergencesless tensors of odd rank.
(We assume that functions,
tensors of rank $0$ belong to $L_+$ also.)
Using equation \eqref{distinguishedlifting}
for the distinguished lifting,
consider the following liftings:
      \begin{equation}\label{distinguishedliftings8}
      \hPi^\rh_{\pm}={\hl+\lo-1\over 2\lo-1}\hPrh(\Delta)\pm
      {\lo-\hl\over 2\lo-1}\left(\hPrh(\Delta)\right)^*\,.
      \end{equation}
$\hPi_+$ is self-adjoint map \eqref{distinguishedlifting}
on the whole space of operators.
It is distinguised lifting map of spaces $\D^{(n)}_\lo(\RR^m)$
for even $n$. Respectively anti-slef-adjoint map
$\hPi_- $ is the
distinguished lifting for odd $n$.
One can show that
$\hPi_+ $ is  $\diff$-equivaraint on the subspace
$L_+$, and respectively
$\hPi_-$ is  $\diff$-equivaraint lifting on the subspace
$L_-$.   This can be checked directly
by applying equations \eqref{distinguished4}.
Namely notice that the canonical lifting
$\hPrh$ (see equation \eqref{canonicalpencillifting1})
is also self-adjoint on $L_+$ and
it is anti-self-adjoint on $L_-$.
  E.g. for $\Delta=S^{ikm}\p_i\p_k\p_m$,
$\hPrh(\Delta)=S^{ikm}\p_i\p_k\p_m$ and
$\left(\hPrh(\Delta)\right)^*=
-\p_i \p_k\p_m \left(S^{ikm}\,\,\right)=-S^{ikm}\p_i\p_k\p_m$
since condition \eqref{divergenceless22} is obeyed
(we work in Cartesian coordinates where $\rh=|Dx|$).
Hence the canonical lifting $\hPrh$ is equal to
$\hPi_+ $ on $L_+$, and
$\hPrh$ is equal to
$\hPi_-$ on $L_-$.
It follows from equations \eqref{distinguished4} and
\eqref{distinguishedliftings8}
that infinitesimal variations with respect to the volume form
of the liftings  $\hPi^{\rh}_{+}$
on $L_+$, and of  $\hPi^{\rh}_{-}$ on $L_-$ vanish. Hence due to
 equation \eqref{isittrueingeneral?} the lifting
$\hPi^\rh_{+ }$ is $\diff(M)$-equivariant
on $L_+$, and
respectively $\hPi^\rh_{-}$ is
$\diff(M)$-equivariant on the subspace $L_-$.
\end{example}

\section {Taylor series for operators on algebra of densities and
 self-adjoint liftings.}\label{section5}

\subsection {Taylor series }
Let $\hD$ be an arbitrary operator defined on
the algebra of densities on a manifold $M$
which is provided with a volume form $\rh$.  Pick an arbitrary weight $\lo$.
Consider the restriction $\Delta_{0}=\Delta\big\vert_{\hl=\lo}$ of the operator
$\Delta$ on densities of weight $\lo$ and the canonical lifting
\eqref{canonicalpencillifting1} of this operator,$\wD 0 =\hPrh(\Delta_{0})$.
The operators $\hD$ and $\wD 0 $ coincide
at $\hl=\lo$, hence we have
expansion
      \begin{equation}\label{taylorseries1}
\hD=\wD 0 +(\hl-\lo)\hD_{(1)}\,.
       \end{equation}
 Using this consideration
repeatedly we come to the expansion of this operator as a power series:
\begin{equation*}
 \hD=\hD_{(0)}=\wD 0 +(\hl-\lo)\hD_{(1)}=
 \wD 0 +(\hl-\lo)\wD 1 +(\hl-\lo)^2\hD_{(2)}=\dots
\end{equation*}
         \begin{equation}\label{taylorseries2}
=\sum_{k=0}^p(\hl-\l_0)^k\wD k +(\hl-\lo)^{p+1}\hD_{(p+1)}=
\sum_{k=0}^n(\hl-\l_0)^k\wD k \,,
\end{equation}
where $n$ is the order of the initial operator $\hD$.

Here $\Delta_j$ is the restriction of an
operator $\hD_{(j)}$ to the subspace $\F_{\l_0}(M)$
and $\wD j $ is the canonical pencil lifting
\eqref{canonicalpencillifting1} of $\Delta_j$:
                $$
     \Delta_j=\hD_{(j)}\big\vert_{\hl=\lo},\quad
     \wD j =\hPrh(\Delta_j), \quad j=0,1,2,\dots
                $$
The operators
$\Delta_k$ and $\hD_k$ have order $\leq n-k$ if
$\hD$ has order $\leq n$.

Formula \eqref{taylorseries2} is an invariant
expression for the expansions \eqref{operatorincoordinates0},
\eqref{operatorincoordinates1}.

\subsection {Description of all (anti)-self-adjoint liftings}
  Now we use the Taylor series expansion for
  writing down a formula for all
 (anti)-self-adjoint liftings of an arbitrary operator.

Pick an operator $\Delta$
of order $n$ acting on densities
of weight $\lo$, $\Delta\in \D^{(n)}_\lo(M)$.
We find all
(anti)-self-adjoint liftings of this operator, more precisely
self-adjoint liftings if $n$
is even and anti-self-adjoint liftings if $n$ is odd
(see remark \ref{why operator is self-adjoint?}).

Consider first the case of $\lo=1/2$.
In this case the operator $\Delta^*$ acts on half-densities also,
so a (anti)-self-adjoint-lifting is possible iff $\Delta^*=(-1)^n\Delta$,
where $n$ is the
order of $\Delta$. Let $\hD$ be such a lifting
of $\Delta=\Delta_0$:  $\hD\big\vert_{\hl=\lo}=\Delta_0$ and
$\hD^*=(-1)^n\hD$. Consider the Taylor series expansion \eqref{taylorseries2}
of the operator $\hD$:
$\hD=\sum_{k=0}^n\left(\hl-1/2\right)^k\wD k $.
Note that canonical liftings
preserve self-adjointness (see \eqref{adjointforcanonical2}).
Recalling also that the operator $\hl-1/2$ is anti-self-adjoint:
$(\hl-1/2)^*=-(\hl-1/2)$ we come from this expansion
to the fact that the
operator $\hD$ is self-adjoint (anti-self-adjoint) lifting
of $\Delta=\Delta_0$
if and only if all the operators
$\{\Delta_0,\Delta_2,\dots,\Delta_{2k},\dots\}$
are self-adjoint (anti-self-adjoint) and
all the operators
$\{\Delta_1,\Delta_3,\dots,\Delta_{2k+1},\dots\}$
are anti-self-adjoint (self-adjoint).

Thus we have described all (anti-)self-adjoint liftings
for arbitrary operator acting on half-densities.

\begin{remark}
Recall that
in the previous section we showed that in general there are no regular
self-adjoint $\diff(M)$-lifting maps on the space $\D^{(n)}_{1/2}(M)$.
Here we considered self-adjoint liftings for
individual operators $\Delta\in \D^{(n)}_{1/2}(M)$.
\end{remark}

Now we consider the case of lifting of operators acting
on densities of an arbitrary weight $\lo\not={1\over 2}$.
    Let $\hD$ be a lifting of
$\Delta=\Delta_0\in \D^{(n)}_\lo(M)$, $\lo\not=1/2$.
Due to \eqref{taylorseries1}
             $$
 \hD={\wD 0 } +(\hl-\lo)\hD_{(1)}\,,
             $$
 where as usual  $\wD 0 $ is canonical lifting
of operator $\Delta_{0}=\hD\big\vert_{\hl=\lo}$.
To find conditions on the
operator $\hD_{(1)}$ such that $\hD$ is (anti)-self-adjoint
it is convenient to consider its Taylor series
expansion \eqref{taylorseries2}
around $\lo^\prime={1/2}$:
$\hD={\wD 0 } +(\hl-\lo)\hD_{(1)}=$
  \begin{equation}\label{selfadjoint16}
{\wD 0 } +(\hl-\lo)
\left[\wD 1 + \left(\hl-1/2\right)\wD 2 +
\left(\hl-1/2\right)^2\wD 3 +\dots+
\left(\hl-1/2\right)^{n-1}\wD {n}  \right]
  \end{equation}
As usual here $\Delta_i$ is an operator of order $\leq n-i$ and
$\wD i $ is the canonical lifting of $\Delta_i$:
$\wD i =\hPrh(\Delta_i)$.
Take the adjoint of this expansion, and using the fact that
$(\hl-1/2)$ is anti-self-adjoint we compare the
terms proportional to the powers $(\hl-1/2)^k$.
We see that the condition of (anti)-self-adjointness
for $\hD$, $\hD^*=(-1)^n\hD$
is equivalent to the equations:
\begin{equation*}
            \Delta_k-(-1)^{n-k}\Delta^*_k=
\left(\lo-{1\over 2}\right)
\left(\Delta_{k+1}-(-1)^{n-k}\Delta^*_{k+1}\right),\quad
            (k=0,1,2,\dots n-1)
                  \end{equation*}
for the operators $\Delta_k,\, k=0,1,2\dots$,
the ``coefficients'' in the power series
\eqref{selfadjoint16},
(The last operator $\Delta_n$ is just a function.)
One can parametrise solutions of this
equation in the following way: For a given operator
$\Delta_0\in \D^{(n)}_\lo(M)$ pick arbitrary operators
standing on even places, operators
$\{\Delta_{2},\Delta_{4},\dots\}$
such that $\Delta_{2k}\in \D^{(n-2k)}_\lo$ ($k=1,2,\dots$).
Then the operators $\{\Delta_{1},\Delta_{3},\dots\}$
at the odd places are equal to
                  \begin{equation}\label{selfadjoint17}
                  \Delta_{2k+1}=
{\Delta_{2k}-(-1)^{n}\Delta^*_{2k}\over 2\lo-1}+
                    (2\lo-1)
                  {\Delta_{2k+2}+(-1)^{n}\Delta^*_{2k+2}\over 4}\,,\quad
                  k=0,2,4,\dots
                  \end{equation}
In particular we see that all operators $\Delta_k$ have order $\leq n-k$.

 Note that in the special case if the defining operators
$\{\Delta_{2},\Delta_{4},\dots\}$ vanish then
for all $k\geq 1$ the operators $\Delta_k$ vanish, and
 formula \eqref{selfadjoint16}
reduces to the distinguished map \eqref{distinguishedlifting}.

\begin{remark}The constructions above fixes the one-to-one
correspondence between the space
of all self-adjoint operator pencils passing through given operators
and the space of sequences of operators $\{\Delta_2,\Delta_4,\dots\}$.
Both these spaces are independent of the
volume form, but the correspondence established
by the formulae above depends on the choice of the volume form
which defines the Taylor series expansion \eqref{selfadjoint17}.

\end{remark}

\begin{example} $n=2$. Self-adjoint liftings are
parameterised by just one operator of order $0$, i.e.
a function: $\Delta_{2}=F(x)$.
Pick an arbitrary operator $\Delta=\Delta_0\in \D^{(2)}_\lo$
($\lo\not=1/2$) and consider
                             $$
  \hD=\wD 0 +(\hl-\lo)\left(\wD 1 +(\hl-1/2)F\right)\,,
                             $$
where due to \eqref{selfadjoint17},
$F$ is an arbitrary function,
$\Delta_1={\Delta_0-\Delta_0^*\over 2\lo-1}+(\lo-{1\over 2})F$,
$\wD 0 =\hPrh(\Delta_0)$ and $ \wD 1 =\hPrh(\Delta_1)$.
This expansion gives all operator pencils $\hD$
passing throug given operator $\Delta_0$ (as always
$\{\Delta_\l\}\colon\, \Delta_\l=\hD\big\vert_{\hl=\l}$).
The family of these operators is in one-to-one correspondence
with functions $F$ on $M$.
This correspondence depends on volume form
\footnote
{Compare with expansion
\eqref{selfadjointoperator0} in section \ref{subsection2.1}
where, for weights $\lo\not=0,1$, the space of self-adjoint operators
passing through a given operator was parametrised by the function
$F=\hD(1)$ independent of the choice of volume form
(see also end of section \ref{subsection4.3}).}.
\end{example}

Now consider anti-self-adjoint liftings for third order operators.

\begin{example}  $n=3$.  Pick an arbitrary third order
operator $\Delta_0\in \D^{(3)}_\lo$, $\lo\not=1/2$.
An antiself-adjoint lifting has the form
       $$
       \hD=\wD 0 +(\hl-\lo)\left
(\wD 1 +(\hl-1/2)\wD 2 +(\hl-1/2)^2\wD 3 \right)\,,
                             $$
where $\Delta_2$ is an arbitrary first order operator
acting on half-densities,
and the operators $\Delta_1$ and $\Delta_3$ (which is a function)
are expressed through the operator $\Delta_2$ and the initial operator
$\Delta_0$. Recalling that by \eqref{geometricalmeaningoffirstorderoperator}
$\Delta_2$ can be represented
as $\Delta_2=\L_\A+S(x)$, where $A$ is a vector field and $S(x)$
is a scalar, and using equations \eqref{selfadjoint17}, we come to

           \begin{equation}\label{pencilsforthirdorderoperator}
     \hD=  {\widehat
  {\Delta_0}-\widehat {\Delta_0^{^*}}\over 2}+
  {2\hl-1\over 2\lo-1}
{\widehat  {\Delta_0}+\widehat {\Delta_0^{^*}}\over 2}+
\left({\hl(\hl-1)-\lo(\lo-1)}\right)
\left(\widehat {\L_A}+
  {2\hl-1\over 2\lo-1}S(x)\right)\,,
       \end{equation}
where $\wD 0 $, $\widehat {\Delta_0^{^*}}$
and $\widehat {(\L_A)}$
are the canonical liftings \eqref{canonicalpencillifting1}
of the initial operator $\Delta_0$, its adjoint,
and of the Lie derivative:
            $$
\widehat {\L_\A}=\hPrh( (\L_\A))=
   \hat\L_\A
-(\hl-\lo){\rm div\,}_\rh \A,
            $$
where $\hat\L_\A$ is the Lie derivative on the algebra of densities
(see equation \eqref{liederivative2}).
Equations \eqref{pencilsforthirdorderoperator} describe all
anti-self-adjoint liftings of third order operator in terms
of arbitrary vector field and scalar function. They contain interesting
geometrical data which we will discuss elsewhere.
\end{example}

\section {Strictly regular $\proj$-liftings
and self-adjoint liftings}\label{section6}

In section \ref{section4} we studied
regular $\sdiff_\rh$-lifting maps
and studied the special role of
distinguished (anti)-self-adjoint
lifting \eqref{distinguishedlifting}.
 In the previous section we studied all
self-adjoint pencils passing through
any given operator $\Delta$. We used for this the Taylor
series expansion based on strictly regular $\sdiff_\rh$-equivariant
canonical liftings \eqref{canonicalpencillifting1}.
In section \ref{section4} we considered
(anti-)self-adjoint liftings such that on the whole space $\D^{(n)}_\lo(M)$,
they are regular $\sdiff$-equivariant lifting maps.
In this section we will give a short description of another construction
of self-adjoint liftings. We will construct self-adjoint liftings
which are linear
maps on $\D^{(n)}_\lo$ equivariant with respect to a
smaller algebra,
the algebra of projective transformations.
 For simplicity we consider here the case if $M$ is $d$-dimensional
affine space,  $M=\RR^d$.
The algebra $\proj(\RR^d)$
is a finite dimensional subalgebra of vector fields on $\RR^d$. This algebra
corresponds to the group of projective transformations of
$\RR P^d$ $=$ ${\rm SL\,}(d+1,\RR)$, the group of
linear unimodular transformations
of $\RR^{d+1}$. The algebra $\proj(\RR^d)$ is generated by
translations $\p_i$, linear transformations $x^i\p_k$,
and special projective transformations $x^ix^k\p_k$,
($i,k=1,\dots,d$).
  Its dimension is equal to $d+d^2+d=(d+1)^2-1$.

Pick an arbitrary $\lo$, we will describe the construction
of a strictly regular $\proj(\RR^d)$-equivariant
pencil map on $\D_\lo(\RR^d)$.
This construction is based on the
classical results of Lecomte and Ovsienko of constructing a
 $\proj$-equivariant full symbol map on differential operators
(see \cite{LecomteOvs1} or the book \cite{OvsTab}).

Symbols of differential operators are
contravariant symmetric tensor fields
on $\RR^d$ which can be identified
with functions on the cotangent bundle
$T^*\RR^d$ which are polynomials on the fibres.
In particular  to an arbitrary
operator $\Delta=S^{i_1\dots i_n}\p_{i_1\dots i_n}+\dots$
one can assign its principal symbol, the contravariant
symmetric tensor field
$[\Delta]=S^{i_1\dots i_n}\xi_1\xi_2\dots\xi_n$
($\xi_i$ are the standard coordinates in
the fibres of the cotangent bundle $T^*\RR^d$).
This assignment defines a canonical $\Diff(\RR^d)$-equivariant map
from $\D^{(n)}_\l(M)$ to the space of symbols. However modules
of differential operators and modules of symbols
are not $\diff$-isomorphic.
Lecomte and Ovsienko
constructed isomorphism,
full symbol map, $\sigma_\l$, such that for any given
weight $\l$ it is a $\proj$-equivariant
 isomorphism, between
the space of linear differential operators on
densities, of a given weight $\l$, and
symbols of operators, i.e. the space of contravariant symmetric
tensor fields on $\RR^d$. This $\proj$-equivariant isomorphism
is uniquely defined by the
normalisation condition of preserving the principal symbol of operator:
       \begin{equation}\label{normalisationcondition8}
  \sigma_\l\left(S^{i_1\dots i_n}\p_1\dots\p_n\right)=
        S^{i_1\dots i_n}\xi_1\dots\xi_n+\dots
       \end{equation}
(Here and later we will identify contravariant symmetric tensor fields
with functions polynomial on the fibers)
For an arbitrary operator
$\Delta=\sum_n L^{i_1\dots i_n}\p_{i_1}\dots\p_{i_n}
\in \D_\lo(\RR^d)$
            \begin{equation}\label{fullsymbolmap1}
           \sigma_\l\left(\Delta\right)=
      \sum_n
         \left(
          \sum_{k=0}^n
         c^{(n)}_k\left(\l\right)
        \p_{p_1}\dots \p_{p_k}L^{p_1\dots p_k i_1\dots i_{n-k}}
           \xi_{i_1}\dots\xi_{i_{n-k}}
         \right)\,,
             \end{equation}
where the coefficients $c^{(n)}_k(\l)$ are polynomials in $\l$,
of degree $\leq n-k$,
$c^{(n)}_0=1$
(normalisation condition \eqref{normalisationcondition8}).
Equivariance of the isomorphism $\sigma_\l$ with respect
to the algebra $\proj(\RR^d)$ and
the normalisation conditions uniquely define these polynomials.
Namely consider the equivariance condition
$\L_\X\sigma_\l(\Delta)=\sigma_\l\left(\ad_\X\Delta\right)$.
These conditions are evidently obeyed for translations and
infinitesimal affine transformations (the vector fields $\p_i$ and $x^i\p_k$).
Considering equivariance for the special
projective transformations,  $\X^{(i)}=x^ix^k\p_k$
\footnote{
$\ad_{\X^{(i)}}\Delta=
     \left(\L_{\X^{(i)}}L\right)^{i_1\dots i_n}
       \p_{i_1}\dots {i_n}-
        {\cal P}^{(i)i_1}_{pq}(\lo,n)S^{pq i_2\dots i_{n-1}}
\p_{i_1}\dots i_{n-1}$,
where
         $$
{\cal P}^{(i)i_1}_{pq}(\lo,n)=n
     \left({n-1\over 2}
              \left(
      \delta^i_p\delta^{i_1}_q+
       \delta^i_q\delta^{i_1}_p
               \right)+
              \l\left(d+1\right)
      \delta^i_p\delta^{i_1}_q
              \right)
	$$

}
and applying these conditions to the full symbol map \eqref{fullsymbolmap1}
one comes to recurrent relations which define the polynomials
$c^{(n)}_k(\l)$ via the polynomials $c^{(n-1)}_{k-1}(\lo)$.
One finds that
\begin{equation}\label{generalformulaforcoefficients}
       c_k^{(n)}=(-1)^k{
         \left(\begin{matrix} n\cr
             k\cr\end{matrix}\right)
      \left(\begin{matrix}\l(d+1)+n-1\cr
             k\cr\end{matrix}\right)
             \over
       \left(\begin{matrix}2n-k+d\cr
             k\cr\end{matrix}\right)}\,,
\end{equation}
where $\left(\begin{matrix} a\cr m\cr\end{matrix}\right)$
are the binomial coefficients. (See for detail \cite{LecomteOvs1}
or book \cite{OvsTab}.)

         One can consider the map $Q_\l$,
which is inverse to the full symbol map,
a so called quantisation map (see \cite{LecomteOvs1} and \cite{OvsTab}).
It follows from normalisation condition
\eqref{normalisationcondition8}, equations
\eqref{fullsymbolmap1} and \eqref{generalformulaforcoefficients}
that for the quantisation map
                      $$
  Q_\l\left(L^{i_1\dots i_n}\xi_1\dots\xi_n\right)=
        L^{i_1\dots i_n}\p_1\dots\p_n+\dots=
                  $$
         \begin{equation*}\label{quantisationmap1}
               \sum_{k=0}^n
        \tilde c^{(n)}_k(\l)
        \p_{p_1}\dots \p_{p_k}L^{p_1\dots p_k i_1\dots i_{n-k}}
           \p_{i_1}\dots\p_{i_{n-k}}\,,
          \end{equation*}
where the $\tilde c^{(n)}_k(\hl)$ are polynomials in $\l$ of order $n-k$,
which can be recurrently obtained from polynomials $c^{(n)}_k(\hl)$
in equation \eqref{generalformulaforcoefficients},
  $c^{(n)}_0=1$, $c^{(n)}_1=-c^{(n)}_1,\dots$.

\begin{example}\label{exampleofsecondorderoperatorproj16}
Consider the full symbol map and its inverse on $\D^{(2)}_\l(\RR)$:
Given $\Delta\in \D^{(2)}_\l(\RR)$,
$\Delta=a(x){d^2\over dx^2}+b(x){d\over dx}+c(x)$
              \begin{equation*}
    \sigma_\l(\Delta)=
      \sigma_\l\left(a(x){d^2\over dx^2}\right)+
      \sigma_\l\left(b(x){d\over dx}\right)+
    \sigma_\l\left(c(x)\right)=
            \end{equation*}
              \begin{equation*}
    \left(a(x)\xi^2-{2\l+1\over 2} a_x(x)\xi+
      {\l(2\l+1)\over 3}a_{xx}(x)\right)+
     \left( b(x)\xi-\l b_x(x)\right)+c(x)\,.
            \end{equation*}
Respectively for the quantisation map $Q_\l=\sigma_\l^{-1}$ we have
               \begin{equation*}
    Q_\l(a(x)\xi^2+b(x)\xi+c(x))=
      Q_\l\left(a(x)\xi^2\right)+
      Q_\l\left(b(x)\xi\right)+
    Q_\l\left(c(x)\right)=
            \end{equation*}
              \begin{equation*}
    \left(a(x){d^2\over dx^2}+{2\l+1\over 2} a_x(x){d\over dx}+
      {\l(2\l+1)\over 6}a_{xx}(x)\right)+
     \left( b(x){d\over dx}+\l b_x(x)\right)+c(x)\,.
            \end{equation*}
\end{example}

Now using the $\proj(\RR^n)$-equivariant full symbol map, $\sigma_\l$,
and its inverse, the quantisation map $Q_\l$ we construct
strictly regular $\proj$-equivariant pencil liftings.

 Pick an arbitrary weight $\l=\lo$ and
consider on the $\D_\lo(\RR^n)$
a map
         \begin{equation*}\label{projectivelifting0}
\D_\lo(\RR^n)\ni\Delta\mapsto \hD=\hPi(\Delta)\colon\quad
             \hPi(\Delta)\big\vert_{\hl=\l}=
       Q_\l\left(\sigma_\lo\left(\Delta\right)\right)\,.
         \end{equation*}
With some abuse of language we say that
         \begin{equation}\label{projectivelifting1}
\hPi(\Delta)\big\vert_{\hl=\l}=
       Q_\hl\left(\sigma_\lo\left(\Delta\right)\right)\,.
         \end{equation}
The map $\hPi$ is a composition of $\proj(\RR^d)$-equivariant maps,
$\hPi_{\hl=\lo}=Q_{\lo}\circ\sigma_\lo={\rm id}$ and it preserves the order of operators.
Hence the map $\hPi$ is a strictly regular $\proj$-equivariant lifting of $\D_\lo(\RR^d)$.
This is a unique lifting, by the uniqueness of map \eqref{fullsymbolmap1}.

Let us look in more detail at the structure of the $\proj$-equivariant
 lifting map $\hPi$.  Pick $n$ and consider the restriction of
this strictly regular lifting map
to the subspace of operators of order $\leq n$.

For any operator $\Delta=\Delta_{(0)}\in \D^{(n)}_\lo(\RR^d)$
consider an operator $\Delta_0$ with the same principal symbol,
$[\Delta]=[\Delta_0]$ such
that the value of the full symbol map $\sigma_\lo$ at $\Delta_0$ is
equal to its principal symbol.
               $$
   \Delta_0=\pr_0 (\Delta)=Q_\lo\left([\Delta]\right)\,.
               $$
The lifting map $\hPi_0=\hPi\circ \pr_0$
maps an operator $\Delta$ to an operator
$\hD_0=Q_{\hl}([\Delta])$
such that for the corresponding operator pencil
$\{\Delta_\l\}$ ($\Delta_\l=Q_\l([\Delta])$) all operators $\Delta_\l$
have the same symbol as the operator $\Delta$.  The operator
$\Delta_{(1)}=\Delta-\Delta_0$ is an operator of order $\leq n$,
$\Delta_{(1)} \in \D^{(n-1)}_\lo$.
Applying the same
procedure to the operator $\Delta_{(1)}$ and using
these considerations repeateadly we come to a $\proj$-equivariant
decomposition of operators in $\D^{(n)}_\lo(\RR^n)$,  and the lifting map
\eqref{projectivelifting1} acts  on the $n+1$ components individually:
            \begin{equation}\label{decomposition}
\Delta=\Delta_{(0)}=\Delta_0+\Delta_1+\dots+\Delta_n\,,\quad\,,
\Delta_{(i)}=\Delta_i+\dots+\Delta_n\,,
             \end{equation}
              $$
\hPi=\hPi_0+\hPi_1+\dots+\hPi_n, \quad, \hPi_k=\hPi\circ \pr_k\,,\quad
              $$
where $
   \Delta_i=\pr_i(\Delta)=Q_\lo\left(\left[\Delta_{(i)}\right]\right)$.
The operators $\Delta_i$ belong to $\D^{(n-i)}_\lo(\RR^d)$, and the operator
pencil $\Pi_i(\Delta)=\Pi_i(\Delta_i)$ is a lifting of the operator
$\Delta_i$. All operators of this pencil have the same symbol.
The decomposition \eqref{decomposition} is a tool to describe
all regular and all self-adjoint regular pencil liftings
which are equivariant with respect to the algebra $\proj(\RR^d)$.

 Let $\Pi$ be an arbitrary $\proj$-equivariant regular lifting
of $\D^{(n)}_\lo(\RR^d)$. Then it follows from uniqueness
arguments that
      \begin{equation}\label{regularprojlifting2}
\Pi(\Delta)=\sum_{k=0}^n P_k(\hl)\hPi_k(\Delta),
       \end{equation}
where $P_k(\hl)$ ($k=0,1,2,\dots n$) are arbitrary polynomials
on $\hl$ of order $n-k$ which obey the conditions that
 any polynomial $P_k(\hl)$ has an order $\leq k$ and
     $P_k(\hl)\big\vert_{\hl=\lo}=1$,
since the lifting is regular and $\Pi(\Delta)_{\hl=\lo}=\Delta$.
It follows from these conditions that
           \begin{equation}\label{normalisationcondition32}
   P_0(\Delta)=1, P_1(\hl)=1+c(\hl-\lo), \quad
\hbox {and in general $P_k(\hl)=1+(\hl-\lo)G_{k-1}(\hl)$}\,,
           \end{equation}
     where $G_{k-1}(\hl)$ is an arbitrary polynomial of order $\leq k-1$.
We see that the space of liftings is a $1+\dots+n={n(n+1)\over 2}$-dimensional
affine space. (Compare with the dimension of the space of regular
$\sdiff_\rh$-liftings in Proposition \ref{proposition1}).

  The same uniqueness arguments imply that
  in the decomposition \eqref{decomposition}
all operators $\hD_i=\hPi_i\hl(\Delta_i)$ are self-adjoint or
 anti-self-adjoint: $\left(\hD_i\right)^*=(-1)^{n-i}\hD_i$.
Hence the regular lifting \eqref{regularprojlifting2}
is a self-adjoint $\proj$-equivariant regular
 lifting if $n$ is even  (respectively anti-self-adjoint lifting
if $n$ is odd) in the case that the polynomials $P_k(\hl)$ obey the additional condition
of self-adjointness:
          \begin{equation*}\label{normalisationcondition64}
   \left(P_k(\hl)\right)^*=(-1)^kP_k(\hl).
           \end{equation*}
           These  conditions with
the conditions from  equation \eqref{normalisationcondition32} imply that
            \begin{equation*}\label{regularprojselflifting2}
   P_0(\Delta)=1,
P_1(\hl)={2\hl-1\over 2\lo-1},\,\,\,\,
  P_2(\hl)=1+b\left(\hl\left(\hl-1\right)-
                    \lo\left(\lo-1\right)\right)
           \end{equation*}
and in  general $         P_{2k}(\hl)=1+
             \sum_{r=1}^{k}c_r \left(t^{2r}(\hl)-t^{2r}(\lo)\right)$ and
           \begin{equation}\label{comestolast}
P_{2k+1}(\hl)={t(\hl)\over t(\lo)}
                   \left(
                   1+
             \sum_{r=1}^{k}
             d_r \left(t^{2r}(\hl)-t^{2r}(\lo)\right),
              \right)\,, \quad \left( \lo \neq \frac{1}{2} \right),
         \end{equation}
         where $c_i,d_j$ are constant coefficients,
$t(\hl)=\hl-{1\over 2}$ is anti-self-adjoint
linear polynomial in $\hl$:
 $t^*(\hl)=\left(\hl-{1\over 2}\right)^*
=-\hl+{1\over 2}=-t(\hl)$.
We see that space of liftings is
${n^2-p(n)\over 4}$-dimensional
affine space, here $p(n)=0$ for even $n$ and $p(n)=1$ for odd $n$.
 (Compare with the dimension of the space of self-adjoint regular
$\sdiff_\rh$-liftings, see proposition\eqref{proposition2}.)

\begin{example}  Consider liftings on second order operators acting
on $\RR$. We
already calculated the full symbol map and the quantisation map in
example \ref{exampleofsecondorderoperatorproj16}.
We have that for $\Delta=
a(x){d^2\over dx^2}+b(x){d\over dx}+c(x)$ acting on densities
of weight $\lo$, $\hPi\left(\Delta\right)=Q_\hl\circ \sigma_\lo(\Delta)$.

 Consider the decomposition \eqref{decomposition}.
  The principal symbol of the operator $\Delta$ is equal to $[\Delta]=a\xi^2$,
  hence (see example \ref{exampleofsecondorderoperatorproj16})
     $$
\Delta_0=\pr_0(\Delta)=Q_\lo(a\xi^2)=
  a(x){d^2\over dx^2}+{2\lo+1\over 2}a_x{d\over dx}+
    {\lo(2\lo+1)\over 6}a_{xx}
         $$
and
                   $$
 \hD_0=\hPi_0(\Delta)=\hPi\left(\pr_0\left(\Delta\right)\right)=
 Q_\hl([\Delta])=Q_\hl(a\xi^2)=
  a(x)\p_x^2+{2\hl+1\over 2}a_x\p_x+{\hl(2\hl+1)\over 6}a_{xx}\,.
          $$
(Note that for differential operators
in the algebra of densities on $\RR$ we have partial derivatives
$\p_x$, not $d\over dx$.)  Respectively
     $
\Delta_{(1)}=\Delta-\Delta_0=
     \left(b-{2\lo+1\over 2}a_x\right)d/dx+
    \left(c-{\lo(2\lo+1)\over 6}a_{xx}\right)$,
       $$
   \Delta_1=\pr_1(\Delta)=Q_\lo(\Delta_{(1)})=
   \left(b-{2\lo+1\over 2}a_x\right){d\over dx}+
  \lo\left(b_x-{2\lo+1\over 2}a_{xx}\right)\,,
        $$

          $$
\hD_1=\hPi_1(\Delta_1)=\hPi\left(\pr_1\left(\Delta\right)\right)=
Q_\hl\left(\left[\Delta_{(1)}\right]\right)=
   \left(b-{2\lo+1\over 2}a_x\right)\p_x+
  \hl\left(b_x-{2\lo+1\over 2}a_{xx}\right)\,,
   $$
and $\Delta_2=\Delta-\Delta_0-\Delta_1=c-\lo b_x+{\lo(2\lo+1)\over 3}$,
  $\hD_2=\hPi_2(\Delta)=c-\lo b_x+{\lo(2\lo+1)\over 3}$.

We have the decomposition of the strictly regular pencil:
      $$
\hD=\hPi(\Delta)=\hPi_0(\Delta)+\hPi_1(\Delta)+\hPi_2(\Delta)=\hD_0+\hD_1+\hD_2\,.
       $$
       All regular $\proj$-equivarariant pencil maps on $\D^{(2)}_\lo(\RR)$
according to \eqref{regularprojlifting2}
are of the form
             $$
\hPi(\Delta)=\hD_0+\left(1+k_1(\hl-\lo)\right)\hD_1+
\left(1+k_2(\hl-\lo)+k_3(\hl-\lo)^2\right)\hD_2\,,
             $$
where $k_1,k_2,k_3$ are constants.
This is a $3$-dimensional affine plane of liftings.
All regular self-adjoint $\proj$-equivarariant
pencil maps on $\D^{(2)}_\lo(\RR)$
according to \eqref{comestolast} have the form
             \begin{equation}\label{lineofselfadjointliftings}
             \hPi(\Delta)=\hD_0+\left({2\hl-1\over 2\lo-1}\right)\hD_1+
\left(1+k\left(\hl(\hl-1)-\lo(\lo-1)\right)\right)\hD_2,
             \end{equation}
 where $k$ is a constant.

This is an affine line of liftings.
Compare these liftings with the canonical
self-adjoint lifting \eqref{selfadjointoperator2}, which  in this case has
the following appearance:
            \begin{equation}\label{selfadjointoperatorforline}
  \hPi_{\rm can}(\Delta)=a(x)\p_x^2+a_x(x)\p_x+
       (2\hl-1)\gamma(x)\p_x+\hl\p_x\gamma+\hl(\hl-1)\theta(x)\,,
       \end{equation}
where the upper connection $\gamma$ and the Branse-Dicke function
$\theta$ are equal to
    \begin{equation}\label{geometricaldata8}
\gamma(x)={b(x)-a_x(x)\over 2\lo-1}, \,
     \theta(x)={1\over \lo(\lo-1)}
              \left(
 c(x)-{\lo\left(b_x(x)-a_{xx}(x)\right)
      \over 2\lo-1}
          \right)\,,
 \end{equation}
($\lo\not=0,1/2,1$).  Now substituting  \eqref{geometricaldata8}
into \eqref{lineofselfadjointliftings}
and comparing with \eqref{selfadjointoperatorforline},
 we find
                  \begin{equation}\label{lineofselfadjointliftings2}
             \hPi_\kappa(\Delta)=\hPi_{\rm can}(\Delta)+
             \kappa\left(\hl(\hl-1)-\lo\left(\lo-1\right)\right)
                      \left(\theta(x)-2\gamma_x(x)+{2\over 3}a_{xx}(x)\right)\,,
             \end{equation}
($\kappa=\lo(\lo-1)k-1$).
 The pencil maps $\hPi(\Delta)$ are regular and $\proj$-equivariant, whilst
$\hPi_{\rm can}(\Delta)$ is a regular $\diff$-equivariant map.
Hence the function
                   \begin{equation}\label{schwarzian}
{\cal S}(x)=\theta(x)-2\gamma_x(x)+{2\over 3}a_{xx}(x)
                   \end{equation}
is invariant under projective transformations since both pencils are.
 Its variation under an arbitrary diffeomorphism
gives us  a cocycle which vanishes on projective transformations.
Using the equations of transformation for $\gamma$, $\theta$
and $a$ (see appendix \ref{appendix1})
 we come to the fact that under the diffeomorphism $y=y(x)$,
                 $$
              {\cal S}|Dx|^2\mapsto {\cal S}|Dx|^2-\left({y_{xxx}\over y_x}-
              {3\over 2}\left({y_{xx}\over y_x}\right)^2\right)
               a(x)|Dx|^2\,,
                 $$
and we see that we come to the Schwarzian. The appearance of this ubiquitous
object here may not be too surprising, we shall delay a more detailed
study of this until later.

\end{example}

\appendix

\section{Connections and upper connections on densities,
and Brance-Dicke functions}\label{appendix1}

{\small
In this Appendix we will give a brief recollection of
the geometrical objects which appear in this article.
For more details see \cite{KhVor2}.

A connection $\nabla$ on the algebra of densities
defines
a covariant derivative of densities
with respect to vector fields.
It obeys the natural linearity properties and the Leibnitz rule:
                 \begin{itemize}
\item $\nabla_{\X}\left(\bs_1+\bs_2\right)=
\nabla_{\X}\left(\bs_1\right)+\nabla_{\X}\left(\bs_2\right)$\,,
\item  $ \nabla_{f\X+g\Y}\left(\bs\right)=
f\nabla_{\X}\left(\bs\right)+g\nabla_{\Y}\left(\bs\right)$\,,
\item $\nabla_{\X}\left(\bs_1 \bs_2\right)=
\nabla_{\X}\left(\bs_1\right)\bs_2+\bs_1\nabla_{\X}\left(\bs_2\right)$
 (in particular $\nabla_{\X}\left(f \bs\right)=
(\p_{\X}f)\,\bs +f\nabla_{\X}\left(\bs\right)$),
\end{itemize}
for   arbitrary densities $\bs$, $\bs_1$ and $\bs_2$,
arbitrary vector fields $\X$ and $\Y$, and arbitrary
functions $f$ and $g$.
Here $\p_\X$ is the ordinary derivative of a
function  along a vector field.

\smallskip

Denote by $\nabla_i$ the covariant derivative with
respect to the vector field $\p_i=\p/\p x^i$.
For an arbitrary density $\bs=s(x)|Dx|^\l$ of weight $\l$,
\begin{equation*}%\label{}
    \nabla_i\bs=\left(\p_i s +\l\gamma_i s\right)|Dx|^\l,
\quad \hbox{where $\gamma_i(x) |Dx|=\nabla_i(|Dx|)$}\,.
\end{equation*}
Under a change of local coordinates $x^i=x^i(x')$, the symbol
$\gamma_a$ transforms in the following way:
\begin{equation*}\label{transformationofgenuineconnection}
   \gamma_i=x^{i'}_i \left(\gamma_{i'}+
\p_{i'}\log \left|\det \frac{\p x}{\p x'}\right| \right)=
   x_i^{i'}\gamma_{i'}-x^j_{j'}x^{j'}_{ij}\,.
\end{equation*}
We  use the shorthand notations for partial derivatives:
 $x^{i'}_i= \lder{x^{i'}\!\!}{x^i}$ and
 $x^{i'}_{jk}=\p^2 x^{i'}\!\!/\p x^j \p x^k$.

The connection also defines the divergence of a vector field
         $$
  {\rm div\,}_\nabla\X=\p_iX^i-\gamma_i X^i\,.
         $$
If the connection $\nabla$ is induced by a volume form $\rh$,
$\nabla=\nabla^\rh$  then
${\rm div\,}_{\nabla^\rh}\X={\rm div\,}_\rh\X=\rh^{-1}\p_i(\rh X^i)$.

Let  $S^{ij}$ be a contravariant tensor field. One can consider
a \emph{contravariant derivative} or  an
\emph{upper connection}  $\uppernabla$  on densities
associated
with $S$.
This notion can be defined by axioms similar to those
for a usual connection. In particular
on volume forms (densities of weight $\l=1$),
we   have
\begin{equation}\label{upperconnection1}
         \uppernabla^i\rh=
 \uppernabla^i\bigl(\rho(x) |Dx|\bigr)=
\left(S^{ij}\p_j\rho+\gamma^i\rho\right)|Dx|\,.
\end{equation}

 Given a contravariant tensor field $S^{ij}$, a
connection $\nabla$ (covariant derivative) induces an
upper connection (contravariant derivative)
$\uppernabla$ by the rule   $\uppernabla^i=S^{ij}\nabla_j$.
 If the  tensor field $S^{ij}$
is non-degenerate, the converse  is  also true.
 A non-degenerate contravariant tensor
field $S^{ij}(x)$ induces a one-to-one correspondence between
upper connections and usual connections.

Under a change of coordinates the symbol
$\gamma^i$ for an upper connection \eqref{upperconnection1} transforms
as follows:
\begin{equation}\label{transformationofupperconnection}
      \gamma^{i'}=x^{i^\prime}_{i}
\left(\gamma^i+S^{ij}\p_j
\log \left|\det\der{x'}{x}\right|\right)\,.
\end{equation}

It is worth noting  that the difference of two
connections on volume forms is a
covector field, the difference of two
upper connections on volume forms is
a vector field.  In other words the space of all
connections (upper connections)
is an affine space associated with the linear
space of the covector (vector) fields.

Consider two important examples of connections on volume forms.

\begin{example}
An arbitrary volume form $\rh$
defines a connection $\nabla^{\rh}$ by the formula $\Gamma_i=
\Gamma^{\rh}_i=-\p_i\log \rho(x)$. If $\bs=s(x)|Dx|^\l$
is a density of
weight $\l$ then $$
  \nabla^\rh_\X \bs=\rh^\l\p_\X\left(\rh^{-\l} \bs\right)=
        X^i(\p_i s+\l \Gamma_i s)|Dx|^\l\,.
                $$
  This is a flat connection, i.e. its curvature vanishes:
$F_{ij}=\p_i\gamma_j-\p_j\gamma_i=0$.

\end{example}

\begin{example}
Let $\nabla^{TM}$ be an affine connection   on a
manifold $M$ (i.e., a connection on the tangent bundle).
It defines a connection on volume forms $\nabla=-{\rm Tr\,} \nabla^{TM}$
with            $\gamma_a=-\Gamma^b_{ab}$
where $\Gamma^a_{bc}$ is the Christoffel symbol for $\nabla^{TM}$.
\end{example}

  Finally we shall mention Branse-Dicke functions.
If $\uppernabla$ is an upper connection for the symmetric contravariant tensor field
$S^{ik}$ then an object $\theta$ is a Branse-Dicke function corresponding
to the upper connection $\uppernabla$ if $\theta-2\gamma^i\Gamma_i+S^{ik}\Gamma_i\Gamma_k$
is a scalar function, where $\Gamma_i$ is an arbitrary connection.
(If one changes the connection $\G_i$ to another, $\tilde \Gamma_i$,
then an expression $\theta-2\gamma^i\Gamma_i+S^{ik}\Gamma_i\Gamma_k$ changes
by a scalar).
It is easy to see that if the upper connection $\gamma^i$ corresponds
to a genuine connection $\gamma_i$, $\gamma^i=S^{ik}\gamma_k$,
 then the function $\gamma^i\gamma_i$ is a Branse-Dicke function.

Under changing of coordinates the Branse-Dicke function transforms as follows:
                          $$
                    \theta'=
        \theta+2\gamma^i\p_i \log J+\p_i\log J\,
          S^{ij}\p_j \log J\,,
          \hbox{where  $J=\log\det\left({\p x^\prime\over \p x}\right)$}\,.
             $$

Branse-Dicke functions naturally arise when analysing second order operators.
They play the role of second order connections.
(See section\ref{subsection2.2}, \cite{KhVor2}and also equation
  \eqref{schwarzian}.)

\section {Proof of lemma \ref{lemma1}}\label{appendix2}

Without loss of generality suppose that $F$ is a function on operators
which act on rapidly decreasing functions on
$d$-dimensional affine space $\RR^d$ ($d\geq 3$),
and we take the volume form $\rh=|Dx|$ in chosen Cartesian coordinates $x^i$.

A differential polynomial $F(\Delta)$ which is invariant with respect to
the algebra $\sdiff_\rh(\RR^n)$ of divergencesless vector fields
has to be invariant with respect to its subalgebra  ${\rm saff}(m,\RR)$
of translations and divergenceless linear transformations.
It is the subalgebra
generated by the vector fields
$\p_i$ and $x^i\p_j-{1\over m}\delta^i_j x^k\p_k$ ($i,j,k=1\dots,d$)
       (translations and infinitesimal unimodular transformations).
Due to classical results of invariant theory, the  algebra
of invariant tensors
is generated by $\delta^i_k$ and $\vare^{i_1\dots i_d}$.
Since the dimension $d$ is greater than $3$
all invariant differential polynomials are linear combinations
of the coefficients of operators and traces of their derivatives.
E.g.a differential polynomial  $F$ on second order
operators which is ${\rm saff\,}$-equivariant
has the following appearance:
           \begin{equation}\label{secondorderoperatorapp1}
           F(S^{ik}\p_i\p_k+T^i\p_i+R)=
           a_1S^{ik}\p_i\p_k+a_2\p_i S^{ik}\p_k+a_3\p_i\p_kS^{ik}+
           b_1T^i\p_i+b_2\p_iT^i+cR\,,
           \end{equation}
where $a_1,a_2,a_3,b_1,b_2,c$ are arbitrary constants
\footnote{In the case of dimension $d=2$
the invariant tensor $\vare^{ik}$ allows to consider maps such as
as $F(A^i\p_i+T)=a\vare^{in}\p_iA^{m}\p_m\p_n+b\vare^{mn}\p_mR\p_n$.}.

Now let $F$ be a linear $\sdiff_\rh$-equivariant differential polynomial,
on the space $\D^{(n)}(\RR^m)$:
             \begin{equation*}\label{invariance12}
             \ad_\X F(\Delta)=F(\ad_\X \Delta),
             \end{equation*}
for an arbitrary divergenceless
vector field $\X$, (${\rm div\,}_\rh\X=\p_i X^i=0$.)
We prove that it implies that
\begin{equation}\label{weprovethat}
F(\Delta)=a\Delta+b\Delta^++c\Delta(1)+d\Delta^{+}(1)\,.
\end{equation}
We use induction on the order of operators.
First we check by straightforward calculations that this
is true for $n=1,2$.
Indeed if $n=2$ then $F$ has the
appearance \eqref{secondorderoperatorapp1},
since it is necessarily ${\rm saff\,}(m,\RR)$-equivariant. One can
then proceed by direct calculations.
For this it is useful to use  that
               $$
 \ad_\X (T^i\p_i)=(\L_K T^i)\p_i, \quad {\rm and}\,\,
 \ad_\X(S^{ij}\p_i\p_j)=\left(\L_\X S^{ij}\right)\p_i\p_j-S^{ij}\p_i\p_jK^r\p_r
               $$
for a divergence-less vector field $\X$, ($\p_iX^i=0$.).

  Comparing the expressions for
$F(\ad_\X \Delta)$ and $\ad_\X F(\Delta)$, where $F$ is defined by
\eqref{secondorderoperatorapp1}, we come to the requirement that
$b_1=a_1-a_2$ and $b_2=-a_3$. This implies
equation \eqref{weprovethat}.

The statement that we want to prove for
arbitrary $n$ immediately follows
from the following
observation.
\begin{lemma}\label{appendixlemma}
Let $F$ be a linear $\sdiff_\rh$-equivariant map on $\D^{(n)}(\RR^m)$
which depends only on the principal symbol of operator  $\Delta$. Then
$F$ vanishes: $F=0$ if $n\geq 3$.
\end{lemma}

The result follows from above as given $F$,
a linear $\textrm{sdiff}_\rh$-equivariant map
on $\D^{(n)}(\RR^m)$ for $n\geq 3$, we have that,
by the inductive hypothesis,
the restriction of $F$ to the subspace
$\D^{(n-1)}(\RR^m)$ obeys the condition
\eqref{weprovethat}, i.e. $F=a\Delta+b\Delta^+
c\Delta(1)+d\Delta^+(1)$ on
$\D^{(n-1)}(\RR^m)$. Hence the map
$F'=F-a\Delta+b\Delta^+c\Delta(1)+d\Delta^+(1)$
on $\D^{(n)}(\RR^m)$ depends only on the principal
symbol of the operator $\D$. Due to the lemma
\ref{appendixlemma}, $F'=0$, i.e.
$F$ has the appearance \eqref{weprovethat}.

It remains to prove lemma \eqref{appendixlemma}. Suppose
$F(\Delta)$ on $\D^{(n)}$ depends only on principal symbol of $\Delta$:
             $$
F\left(S^{i_1\dots i_n}\p_{i_1}\dots \p_{i_n}+\dots\right)=
a_0 S^{i_1\dots i_n}\p_{i_1}\dots \p_{i_n}+
a_1 \p_{i_1}S^{i_1 i_2\dots i_n}\p_{i_2}\dots \p_{i_n}+
a_2\p_{i_1}\p_{i_2}S^{i_1 i_2 i_3\dots i_n}\p_{i_3}\dots \p_{i_n} +\ldots
             $$
It suffices to prove that $a_0=0$.
Namely suppose that $a_0=a_1=a_{k-1}=0$ ($k\geq 1$), then
consider the map
            $S^{i_1\dots i_n}\rightarrow a_k\p_{i_1}\dots \p_{i_k}
            S^{i_1\dots i_{k} i_{k+1}\dots i_n}$.
If $a_1=\dots=a_{k-1}=0$ then this is a map
from the principal symbol of $\Delta$ to the principal symbol of the operator
$F(\Delta)$. Hence this is $\sdiff_\rh$-equivariant map.
On the other hand this map is not $s\diff_\rh$-equivariant if $a_k\not=0$ (if $n\geq 2$).
Hence $a_k=0$ also. Thus we have shown that from the fact that $a_0=0$ it follows that $F=0$.

One can  prove that the condition $a_0=0$ is necessary condition
by straightforward calculations, comparing the expressions for
$\ad_\X F(\Delta)$ and $F(\ad_\X\Delta)$.
 These calculations
can be essentially facilitated if we consider the map $F$ on
operators such that $\p_{i_1}S^{i_1 i_2\dots i_n}\p_{i_2}\dots \p_{i_n}$
identically vanishes in a vicinity of given point. We have that in this case
       $
   F(\ad_\X \Delta)=
F\left((\L_\X S)^{i_1\dots \_n}\p_{i_1}\dots\p_{i_n}\right)=
       $
      $$
  a_0
 \left(\L_\X S\right)^{i_1\dots i_n}\p_{i_1}\dots\p_{i_n}+
  a_1\p_p\left(\L_\X S\right)^{p i_2\dots i_n}\p_{i_2}\dots\p_{i_n}+
  a_2\p_p\p_q\left(\L_\X S\right)^{pq i_3\dots i_n}
     \p_{i_3}\dots\p_{i_n}+\dots
          =
             $$
       $$
  a_0{\cal T}_0-a_1(n-1){\cal T}_1-
 a_2\left((n-2){\cal T}_2-{\cal T}_3\right)+\dots
       $$
            and
$$\ad_\X F(\Delta)=
    \ad_\X\left(a_0 S^{i_1\dots i_n}\p_{i_1}\dots\p_{i_n}\right)=
a_0\left({\cal T}_0-{n(n-1)\over 2}
      {\cal T}_1-
    {n(n-1)(n-2)\over 6}
       {\cal T}_2+\dots\right)
              $$
where we denote
    ${\cal T}_0=
    \left(\L_\X S\right)^{i_1\dots i_n}
     \p_{i_1}\dots \p(i_n)$,
  ${\cal T}_1=
   S^{pq i_3\dots i_n}\p_p\p_qK^{i_2}
    \p_{i_2}\dots\p_{i_n}$,

    \noindent
 ${\cal T}_2=S^{pqs i_4\dots i_n}
     \p_p\p_q\p_s K^{i_3}
     \p_{i_3}\dots\p_{i_n}$ and
  ${\cal T}_3=\p_pS^{qs i_3\dots i_n}
     \p_q\p_s\p_s K^{p}
     \p_{i_3}\dots\p_{i_n}$.
Comparing the terms proportional to ${\cal T}_2$ and ${\cal T}_3$
we see that on one hand $a_2={n(n-1)\over 6}a_0$
and on the other hand $a_2=0$. Hence $a_0=0$.

\end{document}